\documentclass[twocolumn]{emulateapj}

\usepackage{color}
\usepackage{amsmath}
\usepackage{amssymb}

\renewcommand{\vec}[1]{\mbox{\boldmath$#1$}}
\newcommand{\ep}{{\ell^\prime}}
\newcommand{\PP}{\mathbf{P}}
\newcommand{\FF}{\vec{F}}
\newcommand{\DR}{\nabla_\mathrm{Reflux}\cdot}

\newcommand{\ds}{\displaystyle}

\shorttitle{Castro Radiation}
\shortauthors{Zhang et al.}

\begin{document}

\title{CASTRO: A New Compressible Astrophysical Solver. II. Gray
  Radiation Hydrodynamics}

\author{W.~Zhang\altaffilmark{1},
L.~Howell\altaffilmark{2},
A.~Almgren\altaffilmark{1},
A.~Burrows\altaffilmark{3},
J.~Bell\altaffilmark{1}}

\altaffiltext{1}{Center for Computational Sciences and Engineering,
                 Lawrence Berkeley National Laboratory,
                 Berkeley, CA 94720}

\altaffiltext{2}{Center for Applied Scientific Computing,
                 Lawrence Livermore National Laboratory,
                 Livermore, CA 94550}

\altaffiltext{3}{Dept. of Astrophysical Sciences,
                 Princeton University,
                 Princeton, NJ 08544}

\begin{abstract}
  We describe the development of a flux-limited gray radiation solver
  for the compressible astrophysics code, 
  CASTRO.  CASTRO uses an Eulerian grid with block-structured adaptive
  mesh refinement based on a nested hierarchy of logically-rectangular
  variable-sized grids with simultaneous refinement in both space and
  time.  The gray radiation solver is based on a
  mixed-frame formulation of radiation hydrodynamics.  In our
  approach, the system is split into two parts, one part that couples
  the radiation and fluid in a hyperbolic subsystem, and another
  parabolic part that evolves radiation diffusion and source-sink
  terms. The hyperbolic subsystem is solved explicitly with a
  high-order Godunov scheme, whereas the parabolic part is solved
  implicitly with a first-order backward Euler method.

\end{abstract}

\section{Introduction}
\label{sec:intro}

In this paper, we present the development of a gray radiation solver
in our compressible astrophysics 
code, CASTRO.  CASTRO uses an Eulerian grid with block-structured adaptive mesh
refinement (AMR).  Our approach to AMR is based on a nested hierarchy of
logically-rectangular variable-sized grids with simultaneous
refinement in both space and time.  In our previous paper
\citep[][henceforth Paper I]{CASTRO}, we describe our treatment of
hydrodynamics, including gravity and nuclear reactions.  Here, we
describe an algorithm for flux-limited gray radiation hydrodynamics
based on a mixed-frame formulation.

Many astrophysical phenomena involve radiative processes, which often
dominate the energy transport and dynamical behavior of the system.
Some examples include star formation, stellar structure and evolution,
accretion onto compact objects, and supernovae.  Radiation
hydrodynamics simulations are playing an increasingly important role
in modeling these astrophysical systems.

The fundamental equation of radiation transfer is a six-dimensional
integro-differential equation \citep{PomraningBook}, which is
unfortunately very difficult to solve.  Numerical codes typically solve
one or two angular moment equations of the transfer equation.  One common
approach is to solve a two-moment system including radiation energy
density and radiation flux
\citep[e.g.,][]{HayesNorman03,HubenyBurrows07,GonzalezAH07,SekoraStone10}.
The system is closed by an approximate expression for the radiation
pressure.  Another popular approach is to solve the radiation energy
equation only
\citep[e.g.,][]{TurnerStone01,HayesNF06,KrumholzKMB07,SwestyMyra09,CommerconTA11,crash}.
In this approach, the so-called flux-limited diffusion (FLD)
approximation is used for closure \citep{AlmeWilson73}.  The
two-moment approach is more accurate when the radiation is highly
anisotropic and optically-thin, but it is computationally more
expensive than the FLD approach.  However, FLD is a very good
approximation for optically-thick flows.  Furthermore, the FLD
approach can be numerically more robust than the two-moment approach
for systems in the hyperbolic limit.  We have adopted the FLD approach
in the radiation solver of CASTRO. 

CASTRO solves the equations of nonrelativistic radiation
hydrodynamics.  Thus, gas quantities, such as pressure, temperature
and density, are treated as frame-independent because the corrections
are of order $O(v^2/c^2)$, where $v$ is the gas velocity and $c$ is
the speed of light.  However, radiation quantities, such as radiation
energy density, radiation flux, and radiation pressure, in the
comoving frame differ from those in the laboratory frame by order
$O(v/c)$ \citep{MihalasMihalas99}.  Neglecting the $O(v/c)$ terms
potentially leads to erroneous results, especially in the dynamic diffusion
limit where transport of radiation is dominated by motion of the fluid
\citep{Castor72,MihalasKlein82,Castor04}.  For numerical codes, some
authors chose the comoving frame approach in which the radiation
quantities are measured in the comoving frame
\citep[e.g.,][]{TurnerStone01,HayesNorman03,GonzalezAH07,SwestyMyra09,CommerconTA11},
whereas others chose the mixed-frame approach in which the radiation
quantities are computed in the laboratory frame while the opacities
are measured in the comoving frame
\citep[e.g.,][]{MihalasKlein82,HubenyBurrows07,KrumholzKMB07,SekoraStone10}.
A primary weakness of the comoving frame approach is that it does not
conserve energy, whereas the mixed-frame approach is not suitable for
systems in which line transport is important.  In CASTRO, we have
chosen the mixed-frame approach because it conserves the total energy
and is well suited for AMR.

This paper is organized as follows. In \S~\ref{sec:RHD} we present the
governing equations of the mixed-frame gray radiation hydrodynamics
and the mathematical characteristics of the system.  In
\S~\ref{sec:single} we describe the single-level integration scheme.
In \S~\ref{sec:amr} we describe how the integration algorithm is
extended for AMR.  In
\S~\ref{sec:performance} we show the scaling behavior of CASTRO with
radiation.  In 
\S~\ref{sec:tests} we present results from a series of test problems.
Finally, we summarize the results of the paper in \S~\ref{sec:sum}.

\section{Gray Radiation Hydrodynamics}
\label{sec:RHD}

\subsection{Equations of Gray Radiation Hydrodynamics}
\label{sec:eqns}

Assuming local thermodynamic equilibrium, the mixed-frame
frequency-integrated radiation hydrodynamics equations, correct to
$O(v/c)$, can be written as \citep[see e.g.,][]{MihalasKlein82,
  LowrieMH99}:
\begin{align}
  \frac{\partial \rho}{\partial t} + \nabla \cdot (\rho \vec{u}) = { }
  & 0, \\
  \frac{\partial (\rho \vec{u})}{\partial t} + \nabla \cdot (\rho \vec{u}
  \vec{u}) + \nabla p = { } & \frac{1}{c} \chi_{\mathrm{F}}
  \vec{F}_r^{(0)} \nonumber \\
   & - \kappa_{\mathrm{P}}
  (\frac{\vec{u}}{c}) (aT^4 - E_r^{(0)}), \\ 
  \frac{\partial (\rho E)}{\partial t} + \nabla \cdot (\rho E \vec{u} + p
  \vec{u}) = { } & - c \kappa_{\mathrm{P}} ( a T^4 - E_r^{(0)}) \nonumber
  \\
   & + \chi_{\mathrm{F}} (\frac{\vec{u}}{c}) 
  \cdot \vec{F}_{r}^{(0)}, \\  
  \frac{\partial E_r}{\partial t} + \nabla \cdot \vec{F}_r = { } & c
  \kappa_{\mathrm{P}} ( a T^4 - E_r^{(0)}) \nonumber \\
   & - \chi_{\mathrm{F}} (\frac{\vec{u}}{c}) 
  \cdot \vec{F}_{r}^{(0)},  \\  
  \frac{1}{c^2}\frac{\partial \vec{F}_{r}}{\partial t} + \nabla
  \cdot \mathsf{P}_r = { } & -\frac{1}{c} \chi_{\mathrm{F}}
  \vec{F}_r^{(0)} \nonumber \\
   & + \kappa_{\mathrm{P}}
  (\frac{\vec{u}}{c}) (aT^4 - E_r^{0}). \label{eq:radflux}
\end{align}
Here $\rho$, $\vec{u}$, $p$, $T$, and $E$ are the mass density,
velocity, pressure, temperature, and total energy per unit mass
(internal energy, $e$, plus kinetic energy, $u^2/2$), respectively.
$E_r$, $\vec{F}_r$, and $\mathsf{P}_r$ are radiation energy
density, radiation flux, and radiation pressure tensor, respectively.
Note that here the $r$ subscript denotes radiation. The speed of light
and radiation constant are denoted by $c$ and $a$, respectively,
where $a=4\sigma/c$ and $\sigma$ is the Stefan-Boltzmann constant.
$\kappa_{\mathrm{P}}$ and $\chi_{\mathrm{F}}$ are the Planck mean and
flux mean interaction coefficients, both in units of inverse length.
The $(0)$ superscript denotes the comoving frame.
Radiation quantities ($E_r$, $\vec{F}_r$ and $\mathsf{P}_r$)
without the $(0)$ superscript are measured in the lab frame.
Radiation quantities measured in the comoving and lab frames are
related by the Lorentz transformation \citep{MihalasKlein82}. It
should be noted that absorption and scattering coefficients are always
computed in the comoving frame in the mixed-frame approach. Also note
that scattering can be included in the flux mean interaction coefficient.
The whole system is closed by an equation of state for the fluid and a
relation between $\mathsf{P}_r^{(0)}$ and $E_r^{(0)}$,
\begin{equation}
  \mathsf{P}_r^{(0)} = \mathsf{f}^{(0)} E_r^{(0)},
\end{equation}
where $\mathsf{f}^{(0)}$ is the Eddington tensor in the comoving frame.

In the FLD approximation \citep{AlmeWilson73}, the comoving radiation
flux is written in the form of Fick's law of diffusion,
\begin{equation}
    \vec{F}_{r}^{(0)} = - D \nabla E_r^{(0)},
\end{equation}
where the diffusion coefficient $D$ is given by
\begin{equation}
  D = \frac{c\lambda}{\chi_{\mathrm{R}}},
\end{equation}
where $\chi_{\mathrm{R}}$ is the Rosseland mean of the sum of the
absorption and scattering coefficients, and $\lambda$ is the flux
limiter.  We adopt the flux limiter approximation given in
\cite{LevermorePomraning81} as
\begin{align}
  \lambda = { } & \frac{2+R}{6+3R+R^2}, \label{eq:lambda} \\ 
  R = { } & \frac{|\nabla E_r^{(0)}|}{\chi_{\mathrm{R}} E_r^{(0)}}.
  \label{eq:lam-R}
\end{align}
The corresponding radiation pressure tensor is
\citep{Levermore84}
\begin{equation}
  \mathsf{P}_r^{(0)} = \frac{1}{2}[(1-f)\mathsf{I} +
  (3f-1)\hat{\vec{n}}\hat{\vec{n}}] E_r^{(0)},
\end{equation}
where $\mathsf{I}$ is the identity tensor of rank 2, $\hat{\vec{n}}
= \nabla E_r^{(0)} / |\nabla E_r^{(0)}|$, and the Eddington factor $f$
is given by
\begin{equation}
  f = \lambda + \lambda^2 R^2.\label{eq:FLD-f}
\end{equation}
We note that in the optically-thick limit both the flux limiter
$\lambda$ and the Eddington factor $f$ approach $1/3$, whereas in the
optically-thin limit the flux limiter $\lambda$ and the Eddington
factor $f$ approach $0$ and $1$, respectively.

Furthermore we assume that $\chi_{F} = \chi_{\mathrm{R}}$, a common
approximation accurate in optically-thick regions
\citep{MihalasMihalas99}.  Following \citet{KrumholzKMB07}, we keep
terms up to $O(v/c)$, and we drop all terms that are insignificant in
all following regimes: streaming, static diffusion, and dynamic
diffusion limits.  Our radiation hydrodynamics equations now become,
\begin{align}
  \frac{\partial \rho}{\partial t} + \nabla \cdot (\rho \vec{u}) = { } & 0, \label{eq:full-rho}\\
  \frac{\partial (\rho \vec{u})}{\partial t} + \nabla \cdot (\rho \vec{u}
  \vec{u}) + \nabla p + \lambda \nabla E_r = { } & 0, \label{eq:full-rhou}\\
  \frac{\partial (\rho E)}{\partial t} + \nabla \cdot (\rho E \vec{u} + p
  \vec{u}) + \lambda \vec{u} \cdot \nabla E_r = { }
  &  \label{eq:full-rhoE} \\
   - c \kappa_{\mathrm{P}} ( a T^4 & - E_r^{(0)}) , \nonumber \\
  \frac{\partial E_r}{\partial t} + \nabla \cdot \left(\frac{3-f}{2}
    E_r \vec{u} \right) - \lambda \vec{u} \cdot \nabla E_r = { } &
  \label{eq:full-Er} \\
c \kappa_{\mathrm{P}} ( a T^4 - E_r^{(0)}) + \nabla \cdot &
  \left(\frac{c\lambda}{\chi_{\mathrm{R}}} \nabla E_r\right). \nonumber 
\end{align}
The absorption terms on the right hand side of these equations still
include the radiation energy density in the comoving frame, because this
is the the frame in which emission and absorption balance as the material
becomes opaque.  The comoving and lab frame quantities are
related by
\begin{align} 
  E_r^{(0)} = { } & E_r - \frac{2}{c^2} \vec{u} \cdot \vec{F}_r^{(0)}
  + O(v^2/c^2) \\
  = { } & E_r + 2 
  \frac{\lambda}{\chi_{\mathrm{R}}} \frac{\vec{u}}{c}
  \cdot \nabla E_r  + O(v^2/c^2)
\end{align}
in the framework of the FLD approximation. 

\subsection{Mathematical Characteristics of the Hyperbolic Subsystem}
\label{sec:hyper}

Radiation hydrodynamics under the assumption of FLD is a mixed
hyperbolic-parabolic system with stiff source terms.  The equations of
the hyperbolic subsystem are
\begin{align}
  \frac{\partial \rho}{\partial t} + \nabla \cdot (\rho \vec{u}) = { } & 0, \label{eq:hyper-rho}\\
  \frac{\partial (\rho \vec{u})}{\partial t} + \nabla \cdot (\rho \vec{u}
  \vec{u}) + \nabla p + \lambda \nabla E_r = { } & 0, \label{eq:hyper-rhou}\\
  \frac{\partial (\rho E)}{\partial t} + \nabla \cdot (\rho E \vec{u} + p
  \vec{u}) + \lambda \vec{u} \cdot \nabla E_r = { } & 0 , \label{eq:hyper-rhoE}\\
  \frac{\partial E_r}{\partial t} + \nabla \cdot \left(\frac{3-f}{2}
    E_r \vec{u} \right) - \lambda \vec{u} \cdot \nabla E_r = { } &
  0, \label{eq:hyper-Er}
\end{align}
which are obtained by neglecting the terms on the right-hand-side of
Eqs.~\ref{eq:full-rho}--\ref{eq:full-Er}.
In the limit of strong equilibrium (i.e.,
$E_r^{(0)} \approx a T^4$ and $\chi_\mathrm{R} \to \infty$), these
right-hand-side terms are negligible
and the full system becomes hyperbolic, governed by
Eqs.~\ref{eq:hyper-rho}--\ref{eq:hyper-Er}.  In the more general case
the hyperbolic subsystem will be solved first as part of a time-split
discretization.

In CASTRO, we solve the hyperbolic subsystem with a Godunov method,
which utilizes a characteristic-based Riemann solver.  The Godunov
method requires that we
analyze the mathematical characteristics of the hyperbolic subsystem.
For simplicity, let us consider the system in one dimension, which can
be written in terms of primitive variables as,
\begin{equation}
  \frac{\partial{Q}}{\partial{t}} + A \frac{\partial{Q}}{\partial{x}}
  = 0,
\end{equation}
where the primitive variables are
\begin{equation}
  Q = \left( \begin{array}{c}
              \rho \\
              u \\
              p \\
              E_r \end{array}\right),
\end{equation}
and the Jacobian matrix is
\begin{equation}
  A = \left( \begin{array}{cccc}
      u & \rho & 0 & 0 \\
      0 & u    & \frac{1}{\rho} & \frac{\lambda}{\rho} \\
      0 & \gamma p & u & 0 \\
      0 & \frac{3-f}{2}E_r & 0 & (\frac{3-f}{2}-\lambda) u
      \end{array} \right).
\end{equation}
The system is hyperbolic because the Jacobian matrix is diagonalizable
with four real eigenvalues.  In general cases, the eigenvalues and
eigenvectors are unfortunately very complicated.  However,
when the following relation holds,
\begin{equation}
  \frac{3-f}{2} = \lambda + 1, \label{eq:flamb}
\end{equation}
the four eigenvalues are,
\begin{equation} u-c_s, \ u, \ u, \ u+c_s,
\end{equation}   
where 
\begin{equation}
c_s = \sqrt{\gamma \frac{p}{\rho} + (\lambda + 1) \frac{\lambda
    E_r}{\rho}} \label{eq:cs}
\end{equation}
is the radiation modified sound speed.  The corresponding right
eigenvectors are,
\begin{equation}
\left(\begin{array}{c} 1 \\ -c_s/\rho \\ \gamma p/\rho \\
    (\lambda+1)E_r/\rho \end{array}\right), \ 
\left(\begin{array}{c} 0 \\ 0 \\ -\lambda \\ 1 \end{array} \right), \ 
\left(\begin{array}{c} 1 \\ 0 \\ 0 \\ 0 \end{array} \right), \ 
\left(\begin{array}{c} 1 \\ c_s/\rho \\ \gamma p/\rho \\
    (\lambda+1)E_r/\rho \end{array}\right), \label{eq:righteigen}
\end{equation}
and the corresponding left eigenvectors are,
\begin{equation}
  \begin{array}{rrrr}
  (0,& -\rho/2c_s,& 1/2{c_s}^2,& \lambda/2 {c_s}^2 ), \\
  (0,&             0,& -(\lambda+1) E_r / \rho{c_s}^{2},& \gamma  p / \rho {c_s}^2 ), \\
  (1,&             0,& -1 / {c_s}^{2},& -\lambda / {c_s}^2 ), \\
  (0,& \rho / 2c_s,& 1 / 2{c_s}^2,& \lambda / 2{c_s}^2 ).
  \end{array} \label{eq:lefteigen}
\end{equation}
These four eigenvectors define the characteristic fields for the
one-dimensional system.  By computing the product of the right
eigenvectors and the gradients of their corresponding eigenvalues
\citep{LeVeque}, we find that the first and fourth fields are
genuinely nonlinear corresponding to either a shock wave or a
rarefaction wave.  The second and third fields are linearly degenerate
corresponding to a contact discontinuity in either gas pressure or
density.  Note that there is also a jump in radiation energy density
accompanying the jump in gas pressure such that the total pressure,
$p_{\mathrm{tot}} = p + \lambda E_r$, is constant across the contact
discontinuity.  Obviously, in three dimensions, there are two
additional linear discontinuities for transverse velocities just like
the case of pure hydrodynamics.

It should be noted that Eq.~\ref{eq:flamb} is satisfied in both
optically-thick and thin limits.  Although this condition is not always
satisfied, it is a fairly good approximation.  The
\citet{LevermorePomraning81} flux limiter that we use (Eqs.~\ref{eq:lambda}
\& \ref{eq:lam-R}) satisfies
\begin{equation}
  0.978 < \frac{3-f}{2} - \lambda \leq 1.05.
\end{equation}
Thus, for the purpose of an approximate Riemann solver, the
approximate eigenvalues and eigenvectors are used.  

\subsection{Radiation Diffusion and Source-Sink Terms}

The parabolic part of the system consists of the radiation diffusion
and source-sink terms, which were omitted from the discussion of the
hyperbolic subsystem (Eqs.~\ref{eq:hyper-rho}--\ref{eq:hyper-Er}).
\begin{align}
  \frac{\partial (\rho e)}{\partial t} = { } & - c \kappa_{\mathrm{P}} ( a
  T^4 - E_r^{(0)}), \label{eq:drhoe-i0}\\
  = { } & -c \kappa_{\mathrm{P}} (a T^4 - E_{r}) + 2 \lambda
  \frac{\kappa_{\mathrm{P}}}{\chi_{\mathrm{R}}} \vec{u} \cdot 
    \nabla E_r, \label{eq:drhoe-i}\\
  \frac{\partial E_r}{\partial t} = { } & c
  \kappa_{\mathrm{P}} ( a T^4 - E_r^{(0)}) + \nabla \cdot
  \left(\frac{c\lambda}{\chi_{\mathrm{R}}} \nabla E_r\right) \label{eq:dEr-i0}\\
  = { } & c \kappa_{\mathrm{P}} ( a T^4 - E_r) -
  2 \lambda
  \frac{\kappa_{\mathrm{P}}}{\chi_{\mathrm{R}}} \vec{u} \cdot 
    \nabla E_r 
  + \nabla \cdot \left(\frac{c\lambda}{\chi_{\mathrm{R}}} \nabla
    E_r\right), \label{eq:dEr-i}
\end{align}
where $e$ is the specific internal energy.  The term $c
\kappa_{\mathrm{P}} (a T^4 - E_r^{(0)})$ represents the energy
exchange in the comoving frame between the material and radiation
through absorption and emission of radiation.  The term $2 \lambda
(\kappa_{\mathrm{P}}/\chi_{\mathrm{R}}) \vec{u} \cdot \nabla E_r$
is due to the Lorentz transformation of radiation energy density.  We
do not directly solve Eqs.~\ref{eq:drhoe-i0} \& \ref{eq:dEr-i0}
because of our mixed-frame approach.  Instead, we solve
Eqs.~\ref{eq:drhoe-i} \& \ref{eq:dEr-i}.  An implicit treatment is
usually necessary in order to solve these equations because of their
stiffness.  However, the Lorentz transformation term $2 \lambda
(\kappa_{\mathrm{P}}/\chi_{\mathrm{R}}) \vec{u} \cdot \nabla E_r$
can be treated explicitly in many situations because it is of similar
order to the term $\lambda \vec{u} \cdot \nabla E_r$ in the hyperbolic
subsystem (Eqs.~\ref{eq:hyper-rhoE} \& \ref{eq:hyper-Er}), unless the
Planck mean is much larger than the Rosseland mean.  Without
scattering, the Planck mean is usually larger than the Rosseland mean
because the latter gives more weight to the lower opacity part of the
radiation spectrum.  However, in many astrophysical phenomena (e.g.,
shock breakout in core-collapse supernovae), electron scattering,
which contributes to the Rosseland mean only, is the dominant source
of opacity, and therefore the Lorentz transformation term can be
neglected in those cases.

\section{Single-level Integration Algorithm}
\label{sec:single}

For each step at a single level of refinement, the state is first
evolved using an explicit Godunov method for the hyperbolic subsystem
(\S~\ref{sec:explicit}).  Then an implicit update for radiation
diffusion and source-sink terms is performed
(\S~\ref{sec:implicit}).

It is customary in time-split schemes to denote intermediate quantities
with a fractional time index such as $n+1/2$, so that, for example,
the explicit hyperbolic update would advance radiation energy density
from $E_r^n$ to $E_r^{n+1/2}$ and the implicit update would then advance
it from $E_r^{n+1/2}$ to $E_r^{n+1}$.  We are not using this notational
convention here mainly to avoid confusion in the following section with
time-centered quantities constructed at the actual intermediate time
$t^{n+1/2}$.  In \S~\ref{sec:implicit}, where we write out the implicit update in
detail, we will refer to the post-hyperbolic intermediate quantities
as $E_r^-$, $(\rho e)^-$, etc.

\subsection{Explicit Solver for Hyperbolic subsystem}
\label{sec:explicit}

The hyperbolic subsystem is treated explicitly.  This explicit part of
our numerical integration algorithm for radiation hydrodynamics is
very similar to the hydrodynamics algorithm presented in Paper I of this
series.  We refer the reader to Paper I for detailed description
of the integration scheme, which supports a general equation of
state, self-gravity, and nuclear reactions.  Here we will only present
the parts specific to radiation hydrodynamics.

The advection part of the time evolution can be written in the form
\begin{equation}
  \frac{\partial \vec{U}}{\partial{t}} = - \nabla \cdot \vec{F},
\end{equation}
where $\vec{U} = (\rho, \rho \vec{u}, \rho E, E_r)^{T}$ with
the superscript $T$ denoting the transpose operation are the
conserved variables, and $\vec{F}$ is their flux.  The conserved
variables are defined at cell centers.  We predict the primitive
variables, including $\rho$, $\vec{u}$, $p$, $\rho e$, $E_r$, from
cell centers at time $t^n$ to edges at time $t^{n+1/2}$ and use an
approximate Riemann solver to construct fluxes, $\vec{F}^{n+1/2}$,
on cell faces.  This algorithm is formally second-order in both
space and time.  The time step is computed using the standard CFL
condition for explicit methods, with additional constraints if
additional physics (such as burning) is included.  The sound speed
used in the computation is now the radiation modified sound speed $c_s$
(Eq.~\ref{eq:cs}).

\subsubsection{Construction of Fluxes}

CASTRO solves the hyperbolic subsystem of radiation hydrodynamics with
an unsplit piecewise parabolic method (PPM) with characteristic
tracing and full corner coupling \citep{unsplitPPM}.  The four major
steps in the construction of the face-centered fluxes,
$\vec{F}^{n+1/2}$, in the case of hydrodynamics have been described
in details in Paper I.  The extension to radiation hydrodynamics is
straightforward given the characteristic analysis presented in
\S~\ref{sec:hyper}.  Thus, we will not repeat the details
here.  We will also omit the procedures for passively advected
quantities, auxiliary variables, gravity, and reaction, because they
do not change.

First, we compute the primitive variables defined as $\vec{Q}
\equiv (\rho, \vec{u}, p, \rho e, E_r,
p_{\mathrm{tot}}, \rho e + E_r)^T$.  Note that several variables are
redundant for various reasons (i.e., efficiency and safety), which
will be explained later.

Second, we reconstruct parabolic profiles of the primitive variables
within each cell.  The total pressure, $p_{\mathrm{tot}}$, rather than
the gas pressure, $p$, is used in computing a flattening coefficient
\citep{unsplitPPM}.

In the third step, we perform characteristic extrapolations of the
primitive variables and obtain the edge values of $\vec{Q}$ at
$t^{n+1/2}$ using the eigenvectors of the system
(Eq.~\ref{eq:righteigen} \& \ref{eq:lefteigen}) and the parabolic
profiles of the primitive variables.  Flattening is applied in this
procedure.

Finally, the fluxes are computed for the edge values obtained in the
last step using an approximate Riemann solver, which is based on the
Riemann solver of \citet{bellcolellatrangenstein} and
\citet{ColellaGF97}.  The computational procedure is essentially the
same as that in Paper I except that the gas internal energy density,
$\rho e$, and gas pressure, $p$ are now replaced by the total internal
energy density, $\rho e + E_r$, and total pressure, $p + \lambda E_r$.
The Riemann solver computes the Godunov state at the interface, which
is then used to compute the fluxes, $(\rho \vec{u}, \rho
\vec{u}\vec{u} + p \mathsf{I}, \rho E \vec{u}
+ p \vec{u}, ((3-f)/2) E_r \vec{u})^T$.  Recall that there are
redundant variables in the primitive variables, $\vec{Q}$, for
efficiency and safety.  With the total internal energy density, $\rho
e + E_r$, a call to the equation of state can be avoided.  Negative
radiation energy density and gas pressure can also be avoided in the
Godunov state.  The term, $\lambda \vec{u} \cdot \nabla E_r$, in
Eqs.~\ref{eq:hyper-rhoE} \& \ref{eq:hyper-Er} is computed as follows,
\begin{align} 
 (\lambda & \vec{u} \cdot \nabla E_r)_{i,j,k} = \lambda_{i,j,k} \\ 
\times \Big{[} & (\frac{u_{x,i-1/2,j,k} + u_{x,i+1/2,j,k}}{2}) 
           (\frac{E_{r,i+1/2,j,k} - E_{r,i-1/2,j,k}}{\Delta x}) \nonumber \\ 
    + { } & (\frac{u_{y,i,j-1/2,k} + u_{y,i,j+1/2,k}}{2}) 
          (\frac{E_{r,i,j+1/2,k} - E_{r,i,j-1/2,k}}{\Delta y}) \nonumber \\ 
    + { } & (\frac{u_{z,i,j,k-1/2} + u_{z,i,j,k+1/2}}{2}) 
         (\frac{E_{r,i,j,k+1/2} - E_{r,i,j,k-1/2}}{\Delta z})
    \Big{]}, \nonumber
\end{align} 
where the variables with half-integer index are the Godunov states.
The term, $\lambda \nabla E_r$, in Eq.~\ref{eq:hyper-rhou} is
computed in a similar way.

Depending upon a switch set by the user, the Lorentz transformation
term, $2 \lambda (\kappa_{\mathrm{P}}/\chi_{\mathrm{R}}) \vec{u}
\cdot \nabla E_r$, may be included in the explicit update.  

\subsection{Implicit Solver for Radiation Diffusion and Source-Sink Terms}
\label{sec:implicit}

The implicit solver evolves the radiation and gas according to
Eqs.~\ref{eq:drhoe-i} \& \ref{eq:dEr-i}.  The algorithm uses a
first-order backward Euler discretization.  We note that the Lorentz
term, $2 \lambda (\kappa_{\mathrm{P}}/\chi_{\mathrm{R}}) \vec{u}
\cdot \nabla E_r$, may or may not be included in the implicit solver.
The advantage of including this term here is that it effectively
balances emission against absorption in the comoving frame as the
material becomes optically-thick and this coupling becomes stiff.
Implicit treatment also helps avoid a time step restriction when
$\kappa_{\mathrm{P}}/\chi_{\mathrm{R}}$ is significantly greater than
1.  The disadvantage is that it makes the resulting linear systems
nonsymmetric, but this not a major concern in practice.  

The radiation update algorithm is based on that of
\citet{HowellGreenough03}.  The update from the post-hyperbolic state to time
$t^{n+1}$ for Eqs.~\ref{eq:drhoe-i} \& \ref{eq:dEr-i} has the form
\begin{align}
\frac{\rho e^{n+1} - \rho e^{-}}{\Delta t} = 
   - { } & c \kappa_{\mathrm{P}}^{n+1} \left[a(T^{n+1})^4 
   - E_r^{n+1}\right] \label{eq:be1} \\ 
   + { } & q^{n+1} \vec{u}\cdot \nabla E_{r}^{n+1}, \nonumber \\
\frac{E_r^{n+1} - E_r^{-}}{\Delta t} =
   + { } & c \kappa_{\mathrm{P}}^{n+1} \left[a(T^{n+1})^4 -
     E_r^{n+1}\right]  \label{eq:be2} \\
   - { } & q^{n+1} \vec{u} \cdot  \nabla E_r^{n+1}
   + \nabla \cdot \left(d^{n+1} \nabla E_r^{n+1} \right), \nonumber
\end{align}
where $q^{n+1} =
2\lambda^{n+1}\kappa_{\mathrm{P}}^{n+1}/\chi_{\mathrm{R}}^{n+1}$ and
$d^{n+1} = c\lambda^{n+1}/\chi_{\mathrm{R}}^{n+1}$.  Here, we use the $-$
superscript to denote the state following the explicit update, and the
$n+1$ superscript for the state at $t^{n+1}$.  Since the velocity does
not change in the implicit radiation update, we have dropped the $-$
superscript for $\vec{u}^{-}$.  We solve
Eqs.~\ref{eq:be1} \& \ref{eq:be2} iteratively via Newton's method.  We
define
\begin{align}
  F_e = \rho e^{n+1} - \rho e^{-} - \Delta t & \left\{ -c
    \kappa_{\mathrm{P}}^{n+1} \left[a(T^{n+1})^4 - E_r^{n+1}\right] \right.
    \nonumber \\
    \ \ \ & + q^{n+1} \vec{u}\cdot \nabla E_{r}^{n+1} \left. \right\}, \\
 F_r = E_r^{n+1} - E_r^{-} 
  - \Delta t & \left\{
      c \kappa_{\mathrm{P}}^{n+1} \left[a(T^{n+1})^4 -
        E_r^{n+1}\right] \right.
    \nonumber \\
    - q^{n+1} \vec{u} \cdot  \nabla E_r^{n+1} 
    & + \nabla \cdot \left(d^{n+1} \nabla E_r^{n+1} \left. \right)
   \right\}.
\end{align}
Here, we have dropped the $-$ superscript for $\rho^{-}$ because the
implicit update does not change the mass density.
The desired solution is for $F_e$ and $F_r$ to both be zero, and the
Newton update to approach this state is the solution to the linear system
\begin{equation}
  \left[\begin{array}{cc}
      ({\partial F_e}/{\partial T})^{(k)}
      & ({\partial F_e}/{\partial E_r})^{(k)} \\[3pt]
      ({\partial F_r}/{\partial T})^{(k)}
      & ({\partial F_r}/{\partial E_r})^{(k)}
      \end{array} \right]
  \left[\begin{array}{c}
      \delta T^{(k+1)}\\
      \delta E_r^{(k+1)}
      \end{array}\right]
= \left[\begin{array}{c}
      - F_e^{(k)}\\
      - F_r^{(k)}
      \end{array}\right].
\end{equation}
Here $\delta T^{(k+1)} = T^{n+1,(k+1)} - T^{n+1,(k)}$ and
$\delta E_r^{(k+1)} = E_r^{n+1,(k+1)} - E_r^{n+1,(k)}$,
where the $(k)$ superscript denotes the stage of the Newton iteration.
To reduce clutter we drop the $n+1$ superscript without loss of
clarity, so $E_r^{(k+1)} \equiv E_r^{n+1,(k+1)}$.

To solve this system we eliminate the dependency
on $\delta T$ by forming
the Schur complement, leaving a modified diffusion equation for the
radiation update $\delta E_r$.  For simplicity we drop the temperature
derivatives of $d$ and $q$, keeping only the temperature dependence of
the emission and absorption terms.  This does not affect the converged
solution and in practice does not appear to significantly degrade the
convergence rate.

After some mathematical manipulation we obtain the following
diffusion equation, which must be solved for each Newton iteration:
\begin{align}
  \left[(1-\eta) c \kappa_{\mathrm{P}}
        + \frac{1}{\Delta t} \right] E_r^{(k+1)}
  - { } & \nabla \cdot \left(d \nabla E_r^{(k+1)} \right)
   \label{eq:diffusion}
 \\ + { } & (1-\eta)q\vec{u} \cdot \nabla E_r^{(k+1)} 
 \nonumber \\ 
 = (1-\eta) c \kappa_{\mathrm{P}}
      a \left(T^{(k)}\right)^4 + \frac{1}{\Delta t} &
        \left[E_r^- - \eta (\rho e^{(k)} -  \rho e^-) \right], \nonumber
\end{align}
where
\begin{equation}
\eta = 1 - \frac{\rho c_v}{\ds \rho c_v + c\Delta t
   \frac{\partial}{\partial T}
   \left[ \kappa_{\mathrm{P}}\left( a T^4 - E_r \right) \right] }
\approx -\frac{\partial F_r}{\partial T} \left(
         \frac{\partial F_e}{\partial T} \right)^{-1},
\end{equation}
and $c_v$ is the specific heat capacity of the matter. 

The iterations are stopped when the maximum of $|\delta
E_r^{(k+1)}/E_r^{n+1,(k+1)}|$ on the computational domain
falls below a preset tolerance (e.g., $10^{-6}$). 
Note that Eq.~\ref{eq:diffusion} is rewritten to be in terms of
$E_r^{n+1,(k+1)}$ rather than $\delta E_r^{(k+1)}$.  This is done
for computational efficiency.  After one or two Newton iterations,
the solution at the previous iteration, $E_r^{n+1,(k)}$, is getting very
close to the final converged solution $E_r^{n+1}$.  Since we use
$E_r^{n+1,(k)}$ as the starting point for the call to the iterative linear
solver these calls get cheaper for each additional Newton iteration.
If we solved for $\delta E_r^{(k+1)}$ instead, we would have to change
the linear solver tolerance to avoid an unnecessarily accurate and
expensive solve for what may be a very small correction.

Each time we solve the diffusion equation for a new iterate
$E_r^{n+1,(k+1)}$, we update the gas internal energy density as follows,
\begin{align}
  \rho e^{(k+1)} & = \eta \rho e^{(k)} + (1-\eta) \rho e^{-} \label{eq:rhoeup} \\
  - \Delta t (1-\eta) & \left[
    c\kappa_{\mathrm{P}} \left( a(T^{(k)})^{4} - E_r^{(k+1)} \right)
    - q \vec{u} \cdot \nabla E_r^{(k+1)}
  \right]. \nonumber
\end{align} 
The new temperature $T^{n+1,(k+1)}$ then derives from $e^{n+1,(k+1)}$
and $\rho$ via a call to the equation of state.  Other quantities may
or may not be updated: the coefficients $\kappa_{\mathrm{P}}$,
$\chi_{\mathrm{R}}$, $\eta$, $d$, and $q$ are also
temperature-dependent and could have been written with a $(k)$
superscript.  The limiter $\lambda$ can be recomputed based on the new
$E_r^{n+1,(k+1)}$.  It is a tradeoff between efficiency and accuracy
whether to recompute some or all of these at every iteration.  Our
default choice is to recompute all of these at each iteration for
better accuracy.  However, updating coefficients can make the implicit
update iteration less stable if the coefficients are not smooth
functions of their inputs, in addition to the extra computational
costs.  But since each stage of the Newton iteration
(Eq.~\ref{eq:diffusion} combined with Eq.~\ref{eq:rhoeup}) is itself a
conservative solution, the implicit algorithm will conserve total
energy to the tolerance of the linear solver (which should not be
confused with the tolerance of the Newton iterations) regardless of
the number of iterations taken or which coefficients are updated.

In CASTRO, the linear system Eq.~\ref{eq:diffusion}
is solved using the {\it hypre} library
\citep{hypre,hypreweb}.  We have developed drivers that work with
systems in the canonical form
\begin{align}
A E_r^{n+1}
- { } & \sum_i \frac{\partial}{\partial x^i} \left(B_i \frac{\partial
      E_r^{n+1}}{\partial x^i} \right)
+ \sum_i \frac{\partial}{\partial x^i} \left( C_i E_r^{n+1} \right)
\nonumber
\\ + { } & \sum_i D_i \frac{\partial E_r^{n+1}}{\partial x^i}
= \mathrm{rhs},\label{eq:canon}
\end{align} 
where $A$ are cell-centered coefficients and $B_i$, $C_i$, and $D_i$ are
centered at cell faces.  For Eq.~\ref{eq:diffusion} the $C_i$ coefficients
are not used.  There is some subtlety in the appropriate averaging of
coefficients from cells to faces; see \citet{HowellGreenough03} for further
discussion.  The same canonical form works for Cartesian, cylindrical, and
spherical coordinates so long as appropriate metric factors are included
in the coefficients and the rhs.

With {\it hypre} we have a choice between two parallel multigrid
solvers: Schaffer multigrid (SMG) and PFMG.  SMG is more robust for
difficult problems with strongly-varying coefficients, but PFMG is
typically more efficient and scalable.  These solvers work for systems
at a uniform grid resolution (that is, systems associated with a
single level of adaptive mesh refinement).  For systems coupling
together more than one refinement level we could use the {\it hypre}
algebraic multigrid (AMG) solver, or an FAC-type scheme
\citep{McCormick89} using structured solvers on the separate levels.
In earlier versions of the AMR algorithm we required multilevel
solvers for conservative coupling between refinement levels.  We have
now developed a scheme, though, that eliminates the need for a
multilevel linear solver while still conserving total energy.  We
discuss this in detail in the following sections.  Note also that use
of the $D_i$ coefficients deriving from the Lorentz term make the
diffusion equation nonsymmetric.  The multigrid solvers mentioned
above are designed for use with symmetric systems, but good
convergence behavior can still be obtained by using these solvers as
preconditioners for a Krylov method such as GMRES.

\section{AMR }
\label{sec:amr}

\begin{figure}
\plotone{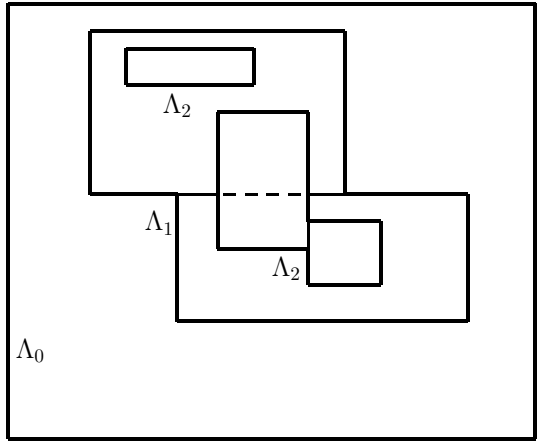}
\caption{A properly nested hierarchy of grids.  Each grid consists of
  a number of cells.  Thick lines represent level boundaries.  The
  union of fine grids at level $\ell$ is contained within the union of
  coarser grids at level $\ell-1$.  
  \label{fig:refspace}}
\end{figure}

\begin{figure}
\plotone{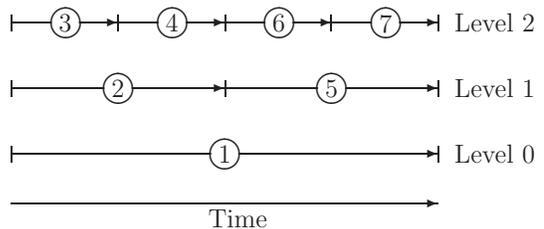}
\caption{One coarse time step for an adaptive run with one base level
  and two refinement levels.  The numbers mark the order of the
  steps. 
\label{fig:reftime}}
\end{figure}

CASTRO uses a nested hierarchy of logically-rectangular,
variable-sized grids with simultaneous refinement in both space and
time, as illustrated in Figures~\ref{fig:refspace} and~\ref{fig:reftime}.
One major design objective of the AMR algorithm is to preserve
the conservation properties of the uniform-grid discretization.
AMR for hyperbolic equations in CASTRO was described in detail
in Paper I.

The explicit update step for radiation hydrodynamics
(\S~\ref{sec:explicit}) follows the same pattern as other hyperbolic
equations and so does not increase the complexity of the AMR
algorithm.  We note that the
hyperbolic subsystem (Eqs.~\ref{eq:hyper-rho} --
\ref{eq:hyper-Er}) is only partially in conservation law form.  It will
not conserve total momentum because it does not include an equation
analogous to Eq.~\ref{eq:radflux} tracking the radiation momentum.  It will
conserve total energy, though, so long as the divergence terms are
differenced in a conservative manner.  These
divergence terms therefore require AMR reflux operations, as described
in \cite{BerCol89}.  The term $\lambda
\vec{u} \cdot \nabla E_r$ appears in both the gas energy and
radiation energy equations with opposite signs, so any consistent
discretization of these terms will conserve total energy.

The AMR version of the implicit radiation diffusion update is
based on \citet{HowellGreenough03},
but with several important differences:  The present algorithm is
fully-implicit, not time-centered.  The optional multilevel linear solve at
the beginning of each coarse time step is no longer included---this feature
was introduced to improve accuracy but we now consider it unnecessary
in most cases.  Finally, the multilevel linear solve for flux
synchronization between coarse and fine levels is replaced by a new
algorithm we call the ``deferred sync.''  These changes entirely
eliminate the need to compute linear system solutions coupling different
levels of the AMR hierarchy, while not compromising conservation of total
energy.  Performance is significantly improved because multilevel linear
solvers tend to be more complex and expensive than those for single-level
systems. 

\subsection{AMR Time Step Outline with Deferred Sync}
\label{sec:dsync}

The AMR time step is defined recursively in terms of operations on a
level $\ell$ and its interactions with coarser and finer levels.
We consider advancing level~$\ell$ from time index $n$ to $n+1$,
corresponding to time values $t^{{\rm old},\ell}$ and $t^{{\rm new},\ell}$,
respectively.
(Even though levels other than $\ell$ have executed different
numbers of time steps, we will
use the $n+1$ superscript to refer to values at time $t^{{\rm new},\ell}$
on all levels involved in the calculation.)
The region covered by level~$\ell$ is denoted $\Lambda^\ell$,
its border is $\partial\Lambda^\ell$, and the border of the next finer
level, projected onto level~$\ell$, is $\PP(\partial\Lambda^{\ell+1})$.

The notation to describe all of this is unavoidably complex due to
the quantities at different
times, levels, and stages of the update process.  In the following
outline, we specify the update for the level $\ell$ and its
synchronization with finer levels.  We include the hyperbolic update
and the refluxing step associated with it in order to show the
proper sequence of operations and to contrast the explicit reflux with
the implicit deferred sync.  As in \S~\ref{sec:explicit} the hyperbolic
conserved state vector is denoted by $\vec{U}$, but we denote the hyperbolic
flux by $\FF_\mathrm{H}$ to distinguish it from the radiation flux.
For the radiation flux we are concerned here only with the diffusion
term in the implicit update, and we denote the associated flux by
$\FF_\mathrm{R}$ to distinguish it from the complete radiation
flux $\FF_r$ introduced in \S~\ref{sec:eqns}.

Note that while the flux divergence is needed everywhere, the fluxes
$\FF_\mathrm{H}$ and $\FF_\mathrm{R}$ themselves are stored only on the
borders between levels.  Our code
has data structures called flux registers designed for this purpose.
The notation $\langle\cdot\rangle$ indicates an average of level $\ell
+ 1$ data in
space over the fine cells (or cell faces) making up each
corresponding coarse cell (or face), while
$\langle\langle\cdot\rangle\rangle$ denotes an average of level $\ell
+ 1$ data in both space and
time over the coarse (level~$\ell$) time step.  Thus, at the only point
in the algorithm where this notation is used,
\begin{equation}
\langle\langle\FF^{\ell+1}\rangle\rangle =
\frac{1}{r^{\ell+1}}\sum_{m=r^{\ell+1}n}^{r^{\ell+1}(n+1)-1}
   \langle \FF^{\ell+1,m+1} \rangle,
\end{equation}
where $r^{\ell+1}$ is the number of level~$\ell+1$ time steps making
up the level~$\ell$ time step from $n$ to $n+1$, and
$\FF^{\ell+1,m+1}$ is the level $\ell+1$ flux during the level $\ell+1$
time step $m+1$.

The expression involving $\DR$ that appears as a deferred source term
in the diffusion equation and explicitly for the hyperbolic reflux is
called the reflux divergence, because it takes the form of the
divergence of a flux difference $\delta\FF$ stored in the flux
registers.  These terms are evaluated {\em only} in the coarse cells
bordering an interface with a finer level.  They do not affect fine
cells at the interface because flux calculations for those cells are
already the more accurate ones; the $\DR$ terms represent the
corrected fluxes from these fine cells being imposed onto the coarse
grid.

Another way to understand the $\DR$ terms
is to consider $\delta\FF_\mathrm{H}$ and $\delta\FF_\mathrm{R}$
to be energy that
has been ``misplaced'' at the coarse-fine interfaces during the level time
step, due to the differing flux calculations on the different levels.  If
the solution were not corrected, this energy would be lost and the system
would not be conservative.  Instead, the $\DR$ terms re-introduce
the missing energy into the system.  For the explicit hyperbolic flux
this is done explicitly; for the radiation flux it contributes to the
right hand side for the implicit update, so as not to impose a new
stability constraint on the size of the time step.

The hyperbolic state vector $\vec{U}$ includes the radiation and
fluid energies updated in the implicit update step, but there is no ambiguity
because the operations and their associated fluxes are
completely distinct.  The refluxing update to $\vec{U}$ is written as
an update, with $\vec{U}$ appearing on both sides of the equation,
because otherwise we would need additional notation to indicate the pre-
and post-reflux states.  Averaging down from fine to coarse levels is also
written as an update.  The meaning of the rest of the pseudocode below
should be reasonably clear in context:

\begin{tabbing}
\quad \=\quad \=\quad \kill
{\bf If} $(\ell < \ell_{\max})$ and
 (regrid requested from base level $\ell$) \\ {\bf then} \\[2pt]
\> {\bf For} $\ep\in\{\ell_{\max}-1,\ldots,\ell\}$ {\bf do} \\[2pt]
\> \> $\bullet$ Determine new grid layout for level $\ep+1$. \\[2pt]
\> \> $\bullet$ Interpolate data to new grids from level $\ep$. \\[2pt]
\> \> $\bullet$ Copy data on intersection with old level $\ep+1$. \\[2pt]
\> {\bf Enddo} \\[2pt]
{\bf Endif} \\[6pt]
{\bf Level Time Step,} level $\ell$: \\[2pt]
\> {\bf Explicit Hyperbolic Update} for $\vec{U}^{\ell,n+1}$\\
   \` (see Paper I) \\[2pt]
\> $\bullet$ Set $\FF_\mathrm{H}^{\ell,n+1}$ for hyperbolic fluxes\\
 \` on ($\PP(\partial\Lambda^{\ell+1})$, $\ell < \ell_{\max}$) and
on ($\partial\Lambda^{\ell}$, $\ell > 0$) \\[2pt] 
\> {\bf Implicit Diffusion Update} \\[2pt]
\> \> $\star$ $\ds \frac{E_r^{n+1} - E_r^{-}}{\Delta t^{n+1}} \; =$ \= $\,
   + c \kappa_{\mathrm{P}}^{n+1} \left[a(T^{n+1})^4 - E_r^{n+1}\right]$ \\[2pt]
\> \> \> $ \mbox{} - q^{n+1} \vec{u} \cdot  \nabla E_r^{n+1}$ \\[2pt]
\> \> \> $ \mbox{} + \nabla \cdot \left(d^{n+1} \nabla E_r^{n+1}
\right)$ \\[2pt]
\> \> \> $ \mbox{} - \left( \Delta t^n / \Delta t^{n+1} \right)
   \DR(\delta\FF_\mathrm{R}^{\ell+1,n})$ \\
   \` on $\Lambda^\ell$ \\[4pt]
\> \> $\star$ $\ds \frac{\rho e^{n+1} - \rho e^{-}}{\Delta t^{n+1}} = 
   - c \kappa_{\mathrm{P}}^{n+1} \left[a(T^{n+1})^4 -
     E_r^{n+1}\right]$ \\[2pt]
\> \> \> $ \mbox{} + q^{n+1} \vec{u}\cdot \nabla E_{r}^{n+1}$
   \` on $\Lambda^\ell$ \\[2pt]
\> {\bf End Implicit Diffusion Update} \\[2pt]
\> $\bullet$ $\FF_\mathrm{R}^{\ell,n+1} =
   -d^{n+1} \nabla E_r^{n+1}$ \\
\` on ($\PP(\partial\Lambda^{\ell+1})$, $\ell < \ell_{\max}$)
   and on ($\partial\Lambda^{\ell}$, $\ell > 0$) \\[2pt]
\> $\bullet$ Advance levels $\ell+1, \ldots,\ell_{\max}$ recursively. \\[2pt]
\> $\bullet$ $\delta\FF_\mathrm{H}^{\ell+1,n+1} =
   \langle\langle\FF_\mathrm{H}^{\ell+1}\rangle\rangle
   - \FF_\mathrm{H}^{\ell,n+1}$ \\
   \` on $\PP(\partial\Lambda^{\ell+1})$, $\ell < \ell_{\max}$ \\[2pt]
\> $\bullet$ $\delta\FF_\mathrm{R}^{\ell+1,n+1} =
   \langle\langle\FF_\mathrm{R}^{\ell+1}\rangle\rangle
   -\FF_\mathrm{R}^{\ell,n+1}$ \\
   \` on $\PP(\partial\Lambda^{\ell+1})$, $\ell < \ell_{\max}$ \\[2pt]
{\bf End Level Time Step} \\[6pt]
{\bf If} $(\ell < \ell_{\max})$ {\bf then} \\
\`  (synchronization/refluxing between levels $\ell$ and $\ell+1$) \\[2pt]
\> $\bullet$ $\vec{U}^{\ell,n+1} := \vec{U}^{\ell,n+1}
   - \left(\Delta t^{n+1}\right)
   \DR(\delta\FF_\mathrm{H}^{\ell+1,n+1})$ \\
 \` on $\Lambda^\ell - \PP(\Lambda^{\ell+1})$ \\[2pt]
\> $\bullet$ $\vec{U}^{\ell,n+1} := \langle\vec{U}^{\ell+1,n+1}\rangle$
      \` on $\PP(\Lambda^{\ell+1})$ \\[2pt]
{\bf Endif}  \` (end synchronization/refluxing).
\end{tabbing}

The advantages of the deferred sync algorithm are that it eliminates the
need for a separate linear solve for synchronization and eliminates the
need for solving multilevel linear systems entirely.  It does, however,
add complications of its own to the
AMR implementation.  One is that there is no longer any point within the
time step cycle when the field variables are fully synchronized.  At the
end of a
coarse time step all levels have reached the same point in time, but if
we want to actually compute the energy budget and confirm conservation
we have to include the contributions from deferred fluxes stored in
flux registers.  These will not be re-introduced into the state until the
next time step.  The end of a coarse time step is also the natural time
for checkpoint/restart operations, so to reproduce the saved state
of the system we now have to include the deferred fluxes in checkpoints
as well.

Regridding becomes an issue as well.  The adaptive algorithm periodically
re-evaluates the refinement criteria for each level and may change
the layout of refined grids.  For grids at level $\ell$ the criteria
are evaluated at level $\ell-1$, and this happens between
level $\ell-1$ time steps.  The interpolation operations between coarse
and fine field variables are conservative.  With the deferred sync, though,
there will also be fluxes stored around the edges of level $\ell$ at the
time that level may be changed.  There is no straightforward way to
transform these stored fluxes so that they coincide with the new
mesh layout.  Instead, what we have is an old set of flux registers that
may now overlap with level $\ell$ as well as with level $\ell-1$, and
if the grid layout changes enough may even overlap with other levels
both finer and coarser.  The deferred sync idea still applies, but the
implementation becomes more complicated than the pseudocode above suggests.
Portions of the stored flux
at the interface between levels $\ell$ and $\ell-1$ may be reintroduced
into the level $\ell$ advance or the level $\ell+1$ advance, and so on,
not just into level $\ell-1$.


\section{Parallel Performance}
\label{sec:performance}

CASTRO is implemented within the BoxLib framework for parallel
structured-grid AMR applications (Paper I and references therein).  In
BoxLib, parallelization is based upon either hybrid
OpenMP-MPI or pure MPI.  Because the {\it hypre} library does not fully
support OpenMP yet, we use the pure MPI approach for the radiation
hydrodynamics solver.  For more information on software design and
parallelization, we refer the reader to Paper I.  Here we show the
scaling behavior of the radiation hydrodynamics solver in CASTRO.

\begin{figure}
\plotone{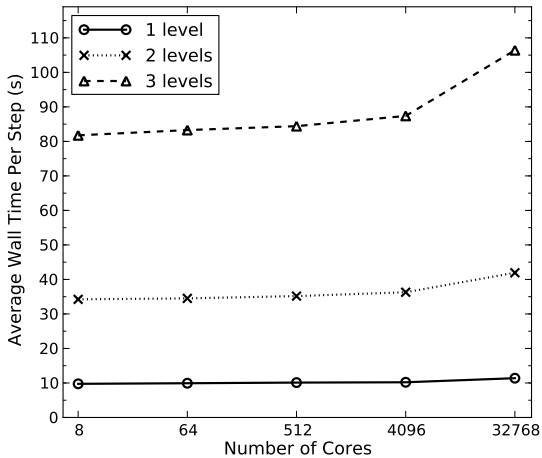}
\caption{Weak scaling behavior of CASTRO on Hopper at NERSC.  Average
  wall clock times per coarse time step are shown for simulations with 1
  ({\it circle}), 2 ({\it X symbol}), and 3 total grid levels ({\it
    triangle}).  The numbers of cells that are advanced in one coarse
  time step increase by a factor of three and seven, for the two- and
  three-level simulations, respectively.
  \label{fig:scaling}}
\end{figure}

A weak scaling study has been carried out on Hopper\footnote{The
  Hopper supercomputer was named after Grace Hopper, a pioneer in
  computer science and the developer of the first compiler.}, a
petascale Cray XE6 supercomputer at the National Energy Research
Scientific Computing Center.  We have performed a series of
three-dimensional simulations with 1, 2 and 3 total levels on various
numbers of cores.  For the convenience of comparison, each run has one
grid of $64^3$ cells at each level on each core.  A refinement factor
of 2 is used in the multi-level simulations.  Thus, the level 1 and 2
grids occupy 12.5\% and 1.5625\% of the whole volume, respectively.  A
point explosion like the one in \S~\ref{test:blast1} is replicated on
each core.  The fine grids are placed at the center of the local
domain of each core.  Figure~\ref{fig:scaling} shows that CASTRO has
very good scaling behavior up to 32768 cores.  For the single-level
simulations, the average wall clock time per coarse time step
increases by only 17\% from 8 cores to 32768 cores.  Because of
subcycling in time for simulations with multiple levels, one coarse
time step consists of one step on the coarse level and two steps on
the next fine level, and the two fine steps might also consist of even
finer steps, recursively.  Thus, the number of cells that are advanced
in one coarse time step increases by a factor of three or seven, for
the two- and three-level simulations, respectively
(Fig.~\ref{fig:reftime}).  Our results also show that the overhead
introduced by AMR is modest.  For example, on 4096 cores, the average
wall clock times per coarse time step for the two- and three-level
simulations are 3.6 and 8.6 times more than that of the single-level
simulation.  This corresponds to an overhead of 19\% and 22\%,
respectively.  On 32768 cores, the average wall clock times per coarse
time step for the two- and three-level simulations are 3.8 and 9.5
times more than that of the single-level simulation.  This corresponds
to an overhead of 25\% and 36\%, respectively.  It should be noted
that single-level simulations with equivalent uniform grids would cost
$\sim 4.5$ and $30$ times more time than the corresponding two- and
three-level simulations we have run even if the single-level
simulations are assumed to scale perfectly.  We also note that in this
series of simulations about half of the time is spent on the implicit
evolution of the parabolic part of the system.

\section{Test Problems}
\label{sec:tests}

In this section we present detailed tests of the code demonstrating
its ability to handle a wide range of radiation hydrodynamics
problems.  Note that not every term is included in every test so that
the algorithms for these terms can be tested separately.  We test the
radiation source-sink term in isolation in the approach to thermal
equilibrium test (\S~\ref{test:equlbrm}).  For a nonequilibrium
Marshak wave problem (\S~\ref{test:marshak}), the simulation involves
the parabolic subsystem only.  In \S~\ref{test:thermalwave}, we assess
the accuracy of the AMR algorithms for radiation diffusion using a
thermal wave test, and perform a convergence study.  The system in a
radiation front test (\S~\ref{test:lightfront}) is in the
optically-thin streaming limit, whereas the system in a shock tube
problem (\S~\ref{test:shocktube}) is in the limit of strong
equilibrium with almost no diffusion.  The full radiation
hydrodynamics system is included in a nonequilibrium radiative shock
problem (\S~\ref{test:2Tshock}) and the advection of a radiation pulse
problem (\S~\ref{test:pulse}).  In \S~\ref{test:hse}, we demonstrate
the ability of CASTRO to maintain a static equilibrium of the gas and
radiation pressures.  In a radiative blast wave test
(\S~\ref{test:blast1}), we compare the results of simulations in 1D
spherical, 2D cylindrical ($r$ and $z$), and 3D Cartesian coordinates.
Finally, we demonstrate the ability of CASTRO to handle a large
Lorentz transformation term in another radiative blast wave test
(\S~\ref{test:blast2}).

A CFL number of 0.8 is used for these tests unless stated otherwise or
a fixed time step is used.  The refinement factor is 2 for all AMR
runs.  The relative tolerance for the Newton iterations in the
implicit update is $10^{-6}$ for all runs.  The Lorentz transformation
term is handled explicitly except in the second radiative blast wave
test (\S~\ref{test:blast2}).

\subsection{Approach to Thermal Equilibrium}
\label{test:equlbrm}

\begin{figure}
\plotone{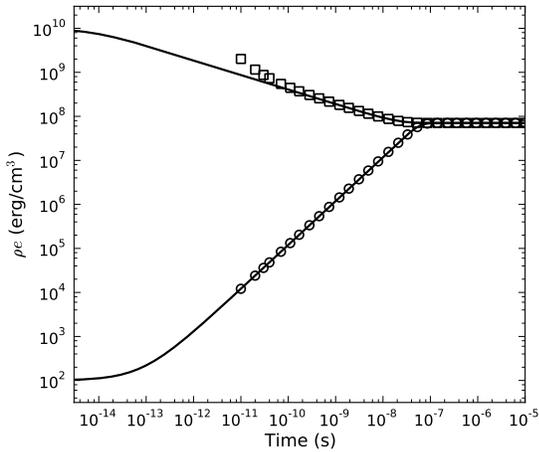}
\caption{ Evolution of internal energy density of gas for calculations
  of the approach to thermal equilibrium.  Numerical results are shown in
  symbols, whereas the analytic solutions are shown in solid lines.
  Two cases with an initial internal energy density of
  $10^{10}\,\mathrm{erg\:cm}^{-3}$ (upper solid line and squares) and
  $10^{2}\,\mathrm{erg\:cm}^{-3}$ (lower solid line and circles) are
  shown. The time step is fixed at $\Delta t = 10^{-11}\,\mathrm{s}$. 
  \label{fig:equlbrm}}
\end{figure}

This test introduced by \citet{TurnerStone01} has often been used to
test the ability of a code to handle the source-sink term, $c
\kappa_{\mathrm{P}} (a T^4 - E_{r})$.  The test consists of a static
uniform field of gas and radiation.  The gas has a density of $\rho =
10^{-7}\,\mathrm{g\:cm}^{-3}$, a Planck mean absorption coefficient of
$\kappa_{\mathrm{P}} = 4 \times 10^{-8}\,\mathrm{cm}^{-1}$, a mean
molecular weight of $\mu = 0.6$, and an adiabatic index of $\gamma =
5/3$.  The initial radiation energy density is $E_r =
10^{12}\,\mathrm{erg\:cm}^{-3}$ corresponding to a temperature of
$\sim 3.39 \times 10^{6}\,\mathrm{K}$. Two cases with an initial
internal energy density of $10^{2}\,\mathrm{erg\:cm}^{-3}$ and
$10^{10}\,\mathrm{erg\:cm}^{-3}$, respectively, have been studied.
The gas temperatures are $\sim 4.81\,\mathrm{K}$ and $\sim 4.81 \times
10^{8}\,\mathrm{K}$ for the two cases, respectively.  The time step is
chosen as $\Delta t = 10^{-11}\,\mathrm{s}$.  The evolution of the
system will bring the gas and radiation into a thermal equilibrium.
The radiation energy density will hardly change because the energy
exchange between the gas and radiation is only a small fraction of the
radiation energy.  Therefore an analytic solution can be calculated by
solving the following ordinary differential equation
\begin{equation}
  \frac{d(\rho e)}{dt} = -c \kappa_{\mathrm{P}} (a T^4 - E_r),
\end{equation}
where $E_r$ is assumed to be constant.  The results of this test of
the approach to thermal equilibrium are shown in
Figure~\ref{fig:equlbrm}.  The agreement with the analytic solution is
good, especially in the first case.  For the second case, the cooling
in the numerical calculation is initially slower than that in the
analytic solution for the first few steps in agreement with the
results of \citet{TurnerStone01}.  The slow cooling is a
  result of the {\em backward} Euler method and a relatively large
  time step at the beginning of the decay of the gas temperature.

\subsection{Nonequilibrium Marshak Wave}
\label{test:marshak}

In this test, we simulate the nonequilibrium Marshak wave problem in
one dimension.  Initially half of the space, $z>0$, consists of a static
uniform zero-temperature gas and no radiation.  A constant radiation
flux $F_{\mathrm{inc}}$ is incident on the surface at $z=0$.  The gas
is not allowed to move in this idealized test.  Thus the gas and
radiation are coupled only through the source-sink term and the system
is governed by Eqs.~\ref{eq:drhoe-i} \& \ref{eq:dEr-i}, with
$\vec{u} = 0$.  \citet{SuOlson96} obtained analytic solutions for
the problem under special assumptions.  The matter is assumed to be
gray with $\kappa_{\mathrm{P}} = \chi_{\mathrm{R}}$, and its volumetric
heat capacity at constant volume is assumed to be $c_v = \alpha T^3$,
where $T$ is the gas temperature and $\alpha$ is a parameter.

\begin{figure}
\plotone{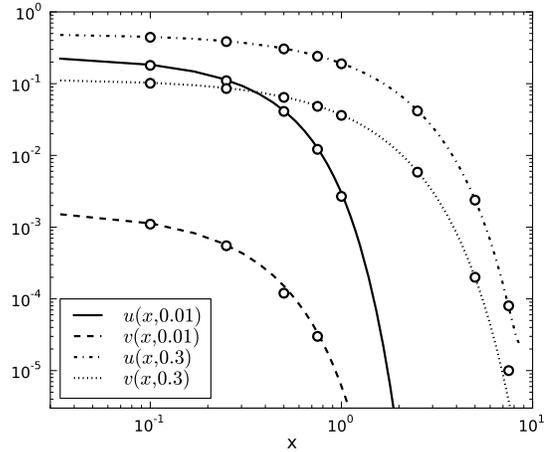}
\caption{ Nonequilibrium Marshak wave.  Numerical results are shown in
  lines, whereas the analytic solutions are shown in circle symbols.
  We show the dimensionless radiation energy density $u$ at $\tau = 0.01$
  ({\it solid line}) and $\tau = 0.3$ ({\it dash-dot line}), and the
  dimensionless gas energy density $v$ at $\tau = 0.01$
  ({\it dashed line}) and $\tau = 0.3$ ({\it dotted line}).  Here
  $\tau$ is the dimensionless time. 
  \label{fig:marshak}}
\end{figure}

We have run this test with $4a/\alpha = 0.1$ and no flux limiter
(i.e., $\lambda = 1/3$).  Figure~\ref{fig:marshak} shows the numerical
results of the dimensionless radiation energy density and gas energy
density defined by \citet{Pomraning79}
\begin{align}
  x \equiv { } & \sqrt{3} \kappa z, \\
  \tau \equiv { } & \left(\frac{4ac\kappa}{\alpha}\right) t, \\
  u(x,\tau) \equiv { } &
  \left(\frac{c}{4}\right)\left(\frac{E_r(z,t)}{F_{\mathrm{inc}}}\right),
  \\
  v(x,\tau) \equiv { } &
  \left(\frac{c}{4}\right)\left(\frac{aT^4(z,t)}{F_{\mathrm{inc}}}\right).
\end{align}
The computational domain of $0<x<5\sqrt{3}$ is covered by 128 uniform
cells.  The dimensionless time step is chosen to be $\Delta \tau = 3
\times 10^{-4}$. The numerical results are in good agreement with the
analytic results of \citet{SuOlson96}.

\subsection{Thermal Wave}
\label{test:thermalwave}

\begin{figure*}
\epsscale{0.85}
\plotone{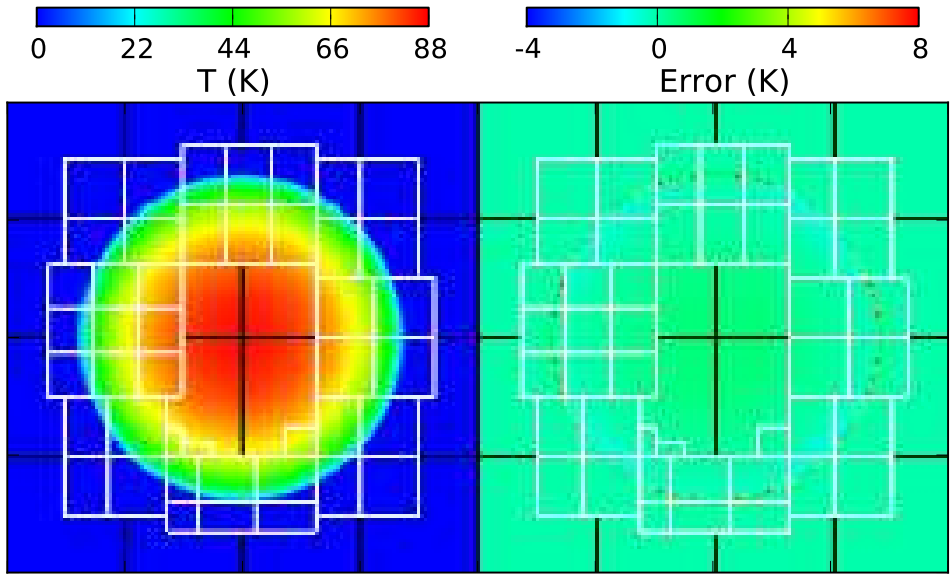}
\caption{ Temperature and error at $t = 0.006\,\mathrm{s}$ for the
  deferred sync run of the thermal
  wave test.  The error is computed as the difference between the
  simulation results and the analytic solution.  We show a 2D slice at
  $x=0$. Also shown is the
  grid structure of the adaptive mesh. The physical domain in each
  dimension is 
  $(-200,200)\,\mathrm{cm}$ for each panel.
  \label{fig:thermalwave}}
\end{figure*}

In this test, we simulate a thermal wave \citep{ZelDovich67} in three
dimensions and use this test to assess the accuracy of the AMR
algorithms for radiation diffusion.  Suppose that a large amount of
energy is deposited into a small volume as the internal energy of
matter.  The matter then starts to radiate and transfer most of its
energy to radiation.  We assume that the Planck mean absorption
coefficient is large enough so that the matter and the radiation are
in thermal equilibrium.  The heat is transported out of the initial
hot spot because of the nonlinear radiation heat conduction.  As a
result, a thermal wave develops.  As the matter cools down, the matter
gains most of the energy again.  Initially the thermal wave speed is
much higher than the sound speed.  Assuming that there is no fluid
motion and the matter contains most of the energy, there is an
analytic solution for this problem \citep{ZelDovich67}.

This test is adapted from \citet{HowellGreenough03}.  The
computational region in this test is three-dimensional with a domain
of $(-200, 200)\,\mathrm{cm}$ in each dimension.  Initially, the
spherical hot spot has an energy of $3 \times 10^{7}\,\mathrm{erg}$
within a radius of $3.125\,\mathrm{cm}$, whereas the ambient medium
has a very low temperature of $10^{-6}\,\mathrm{K}$.  The volumetric
heat capacity is $\rho c_v =
0.05\,\mathrm{erg}\,\mathrm{K}^{-1}\,\mathrm{cm}^{-3}$.  The Planck
mean absorption coefficient is $\kappa_{\mathrm{P}} =
10^{6}\,\mathrm{cm}^{-1}$, whereas the Rosseland mean is
$\chi_{\mathrm{R}} = 10^{-3}
(T/1\,\mathrm{K})^{1/2}\,\mathrm{cm}^{-1}$.  In this test, the
hydrodynamics is turned off, and there is no flux limiter (i.e.,
$\lambda = 1/3$).  We have performed three simulations, a non-AMR run
with $128^3$ uniform cells, an AMR run using the deferred sync
algorithm (\S~\ref{sec:dsync}), and an AMR run using the multilevel
algorithm of \citet{HowellGreenough03}.  The two AMR runs use 2 total
levels with an effective resolution of $128^3$ cells on the finer
level.  A cell is tagged for refinement if its temperature satisfies
both $\nabla T > 0.4 T / \Delta x$ and $T > 10^{-5}\,\mathrm{K}$,
where $\Delta x$ is the size of the cell.  For the non-AMR run, we use
a time step $\Delta t = 1.03^{n-1} \times 5 \times
10^{-16}\,\mathrm{s}$ for step $n$, whereas for the two AMR runs, we
use $\Delta t = 1.0609^{n-1} \times 1.015 \times 10^{-15}\,\mathrm{s}$
for step $n$ on the coarser level.  We have chosen the time steps for
the following reasons.  First, as it expands, the thermal wave is
rapidly decelerated.  Thus, a fixed time step would not be optimal.
Instead, we let the time step grow over time.  Second, the numerical
error depends on both time step $\Delta t$ and cell size $\Delta x$.
In this test, we want to assess the accuracy of the AMR runs in
comparison with a non-AMR run.  Therefore, we make the time step on
the finer level of the AMR runs to be roughly the same as that of the
non-AMR run.  Figure~\ref{fig:thermalwave} shows a 2D slice at $x=0$
of the deferred sync run at $t = 0.006\,\mathrm{s}$ for temperature
and the difference between the numerical and analytic results.  It is
shown that the largest errors ($\sim 5$--$10\,\mathrm{K}$) occur near
the thermal wave front where the slope of the temperature profile is
extremely steep.  At the center, the absolute error is only
$0.6\,\mathrm{K}$, whereas the relative error is 0.7\%. It takes 451
coarse time steps for the AMR runs to reach $t = 0.006\,\mathrm{s}$.
The finer grids in the AMR runs occupy 0.20, 2.3, and 32\% of the
volume at steps 1, 226, and 451, respectively.  Thus, the benefit of
AMR is obvious.  To quantitatively measure the accuracy of the
results, we compute the normalized $L_1$-norm error for temperature,
$\sum_{i,j,k} |T_{\mathrm{n},i,j,k}-T_{\mathrm{a}}(t_\mathrm{n},x_i,y_j,z_k)|
\Delta V_{i,j,k} / \sum_{i,j,k} T_{\mathrm{a}}(t_\mathrm{n},x_i,y_j,z_k)
\Delta V_{i,j,k}$ , where $T_{\mathrm{n},i,j,k}$ and
$T_{\mathrm{a}}(t_\mathrm{n},x_i,y_j,z_k)$ are the numerical and
analytic results
for cell $(i,j,k)$, respectively, and $\Delta V_{i,j,k}$ is the volume
of cell $(i,j,k)$.  At $t = 0.006\,\mathrm{s}$, the $L_1$-norm errors
are $0.00792$, $0.00833$, and $0.00847$, for the non-AMR, deferred
sync, multilevel sync runs, respectively.  The results show that our
AMR algorithms have only increased the error by $5\%$--$7\%$.

\begin{deluxetable}{ccc}
\tablecaption{$L_1$-norm errors and convergence rate for the thermal
    wave test problem.  Four 3D AMR runs with various
    resolutions are shown.  There are two AMR levels in each of
    these runs.
\label{tab:TW}}
\tablewidth{0pt}
\tablehead{
\colhead{Resolution\tablenotemark{a}}    &   \colhead{$L_1$ Error}  &  
\colhead{Convergence Rate} 
}
\startdata
16, 32 & 0.0706 &  \\
32, 64 & 0.0245 & 1.5 \\
64, 128 & 0.00833 & 1.6 \\
128, 256 & 0.00268 & 1.6
\enddata
\tablenotetext{a}{Number of cells across the width of the domain at each of the two AMR levels}
\end{deluxetable}

We have also performed a series of AMR simulations using the deferred
sync to check the convergence behavior of the code.  Besides the AMR
run using the deferred sync that has been presented
(Fig.~\ref{fig:thermalwave}), two lower resolution runs and one higher
resolution run have been performed (Table~\ref{tab:TW}).  In all four
runs, there are two total AMR levels.  In the convergence study, when
the cell size changes by a factor of 2 from one run to another, we
change the time step by a factor of 4.  Table~\ref{tab:TW} shows the
$L_1$-norm errors and convergence rate at $t = 0.006\,\mathrm{s}$.  In
this study, the order of accuracy with respect to $\Delta x$ is $\sim
1.6$.  Because our implicit scheme is first-order in time and
second-order in the spatial discretization of the diffusion term, the
expected convergence rate for a smooth flow is second-order with
respect to $\Delta x$ when $\Delta t / \Delta x^2$ is fixed.  However,
the temperature profile of the thermal wave problem has a very steep
slope near the front and its second derivative is discontinuous there.
Hence, it is not surprising that the achieved order of accuracy is
lower than 2.

\subsection{Optically-Thin Streaming of Radiation Front}
\label{test:lightfront}

\begin{figure}
\plotone{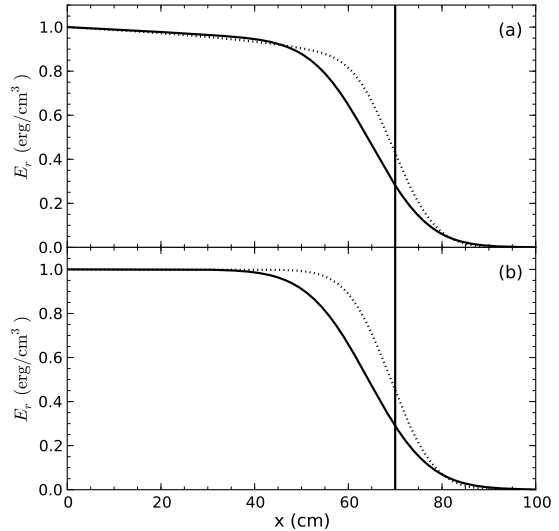}
\caption{ Optically-Thin Streaming of Radiation Front.  We show
  radiation energy density at $t = 2 \times 10^{-9}\,\mathrm{s}$ for
  two cases: (a) $\chi_{\mathrm{R}} = 10^{-4}\,\mathrm{cm}^{-1}$
  (upper panel); (b) $\chi_{\mathrm{R}} = 10^{-7}\,\mathrm{cm}^{-1}$
  (lower panel).  Results with a time step of $\Delta t = \Delta x /
  (2c)$ are shown in solid lines, whereas those with a time step of
  $\Delta t = \Delta x / (20c)$ in dotted lines.  The vertical lines
  at $x = 70\,\mathrm{cm}$ indicate the expected position of the
  radiation front.
  \label{fig:streaming}}
\end{figure}

In this problem, we test our code in the optically-thin streaming
limit.  This is the opposite limit from diffusion.  The computational
domain of this test is a one-dimensional region of $0 < x <
100\,\mathrm{cm}$, covered by 128 uniform cells.  Initially, the
region $x<10\,\mathrm{cm}$ is filled with radiation, $E_r =
1\,\mathrm{erg\:cm}^{-3}$, whereas the region of $x>10\,\mathrm{cm}$
has $E_r = 10^{-10}\,\mathrm{erg\:cm}^{-3}$. The left and right
boundaries are held at $E_r = 1$ and
$10^{-10}\,\mathrm{erg\:cm}^{-3}$, respectively, during the
calculation.  The hyperbolic update (\S~\ref{sec:explicit}) is
switched off in this test.  The Planck mean absorption coefficient is
set to zero. We have studied two cases with Rosseland means of
$\chi_{\mathrm{R}} = 10^{-4}$ and $10^{-7}\,\mathrm{cm}^{-1}$,
respectively.  Thus the optical depths of the whole domain are
$10^{-2}$ and $10^{-5}$, respectively.  For each case, we have
performed two calculations, one with the time step set to $\Delta t =
\Delta x / (2c)$ and one with $\Delta t = \Delta x / (20c)$.
Physically, the radiation front is expected to propagate at the speed
of light, but the diffusion approximation does not naturally support a
propagating front at any speed.  Only the flux limiter permits the
code to approximate the correct solution.  In the numerical results
the radiation front moves at approximately the speed of light with
spreading due to diffusion (Figure~\ref{fig:streaming}).  Better
results are obtained for smaller time steps.

\subsection{Shock Tube Problem In Strong Equilibrium Regime} 
\label{test:shocktube}

\begin{figure}
\plotone{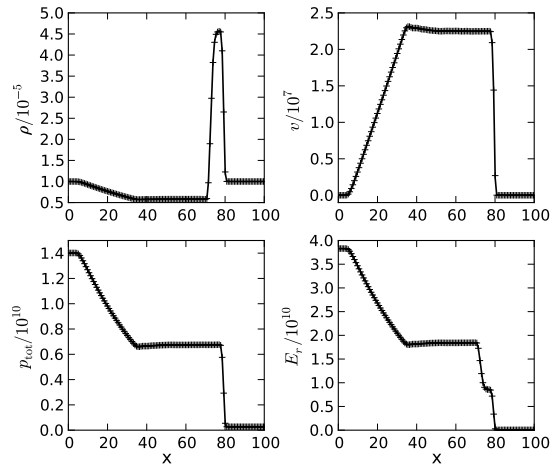}
\caption{ Shock tube at $t = 10^{-6}\,\mathrm{s}$.  Numerical results
  from a full radiation hydrodynamics calculation with 128 cells are
  shown in symbols.  Results from a hydrodynamics calculation with 128
  cells and an EOS for ideal gas plus radiation are shown in
  sold lines for comparison.  We show mass density ($\rho$), velocity
  ($v$), total pressure ($p_{\mathrm{tot}}$), and radiation energy
  density ($E_r$).  The total pressure is the sum of gas pressure and
  radiation pressure.  All quantities are in cgs units.
  \label{fig:shocktube}}
\end{figure}

In this test, the one-dimensional numerical region ($0 < x <
100\,\mathrm{cm}$) initially consists of two constant states: $\rho_L
= 10^{-5}\,\mathrm{g}\:\mathrm{cm}^{-3}$, $T_L = 1.5 \times
10^{6}\,\mathrm{K}$, $u_L = 0$ and $\rho_R =
10^{-5}\,\mathrm{g}\:\mathrm{cm}^{-3}$, $T_R = 3 \times
10^{5}\,\mathrm{K}$, $u_R = 0$, where $L$ stands for the left state,
and $R$ the right state.  The initial discontinuity is at $x =
50\,\mathrm{cm}$.  The gas is assumed to be ideal with an adiabatic
index of $\gamma = 5/3$ and a mean molecular weight of $\mu = 1$.
Initially, the radiation is assumed to be in thermal equilibrium with
the gas (i.e., $E_r = a T^4$).  The interaction coefficients are set to
$\kappa_{\mathrm{P}} = 10^6\,\mathrm{cm}^{-1}$ and $\chi_{\mathrm{R}}
= 10^{8}\,\mathrm{cm}^{-1}$. Thus, due to the huge opacities, the
system is close to the limit of strong equilibrium with no diffusion
(i.e., $E_r \approx a T^4$ and $\chi_\mathrm{R} \to \infty$), and is
essentially governed by Eqs.~\ref{eq:hyper-rho}--\ref{eq:hyper-Er}.
The parameters of this test problem are chosen such that neither
radiation pressure nor gas pressure can be ignored.  The numerical
results are shown in Figure~\ref{fig:shocktube}.  Also shown are the
results from a pure hydrodynamics calculation with an equation of state
(EOS) for ideal gas plus radiation.  As we expected, the results from
the radiation hydrodynamics calculation are almost identical to those
of the pure hydrodynamics calculation.  We also note that our scheme is
stable without using small time steps even though the system is in the
``dynamic diffusion'' limit \citep{MihalasMihalas99, KrumholzKMB07}
with $\tau u/c \sim 10^{7}$, where $\tau$ is the optical depth of the
system.

\subsection{Nonequilibrium Radiative Shock}
\label{test:2Tshock}

Radiation can modify the structure of a shock because it diffuses and
because it interacts with matter.  Analytic estimates of radiative
shock structures can be found in \citet{ZelDovich67,
  MihalasMihalas99}.  Recently \citet{LowrieEdwards08} have found a
semi-analytic exact solution of radiative shocks with gray
nonequilibrium diffusion.  In this section, we present our numerical
results for two shock strengths and compare them to the solutions of
\citet{LowrieEdwards08}.  The first problem is the Mach 2 shock in
\citet{LowrieEdwards08}.  This is a subcritical shock in which the
pre-shock matter is preheated by diffused radiation to a temperature
that is lower than the temperature in the relaxation region far
downstream.  There is an embedded hydrodynamic shock that causes a
jump in density and raises the temperature of the matter above the
final downstream temperature.  The matter temperature behind the
embedded hydrodynamic shock then cools down in the relaxation region.
The second problem is the Mach 5 supercritical shock in
\citet{LowrieEdwards08}.  In the case of a supercritical shock, the
matter temperature just before the hydrodynamic shock equals the
downstream temperature.  Most of the precursor region of a
supercritical shock is close to equilibrium.  Note that in both cases
there is no jump in radiation energy density, otherwise the radiation
flux would be infinite, which is unphysical.

\begin{figure}
\plotone{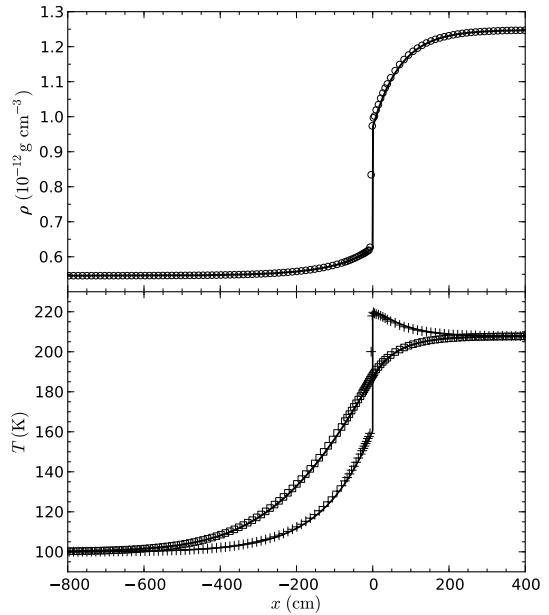}
\caption{ Density and temperature profiles for Mach 2 subcritical
  shock.  Numerical results are shown in symbols.  We show density
  ({\it circle}), gas temperature ({\it plus sign}), and radiation
  temperature ({\it square}).  Here, radiation temperature is defined
  as $(E_r/a)^{1/4}$.  The analytic solution is shown in solid lines.
  We show only part of the region near the hydro-shock.  The
  numerical results have been shifted by $+10\,\mathrm{cm}$
  in space to compensate for the discrepancy in shock position caused
  by the initial setup.
  \label{fig:M2}}
\end{figure}

\begin{figure}
\plotone{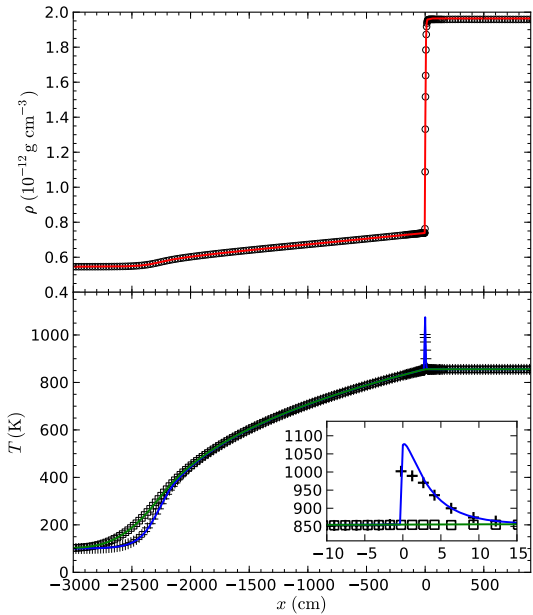}
\caption{ Density and temperature profiles for Mach 5 supercritical
  shock.  Numerical results are shown in symbols.  We show density
  ({\it circle}), gas temperature ({\it plus sign}), and radiation
  temperature ({\it square}).  Here radiation temperature is defined
  as $(E_r/a)^{1/4}$.  Also shown are the analytic results for density
  ({\it red solid line}), gas temperature ({\it blue solid line}), and
  radiation temperature ({\it green solid line}). We show only part of
  the region near the hydro-shock.  The inset shows a blow-up of the
  spike in temperature.  The
  numerical results have been shifted by $-205\,\mathrm{cm}$
  in space to compensate for the discrepancy in shock position caused
  by the initial setup.
  \label{fig:M5}}
\end{figure}

The Mach 2 shock problem is simulated in a one-dimensional 
region of $-1000\,\mathrm{cm} < x < 500\,\mathrm{cm}$ consisting of
two constant states: $\rho_L = 5.45887 \times
10^{-13}\,\mathrm{g}\:\mathrm{cm}^{-3}$, $T_L = 100\,\mathrm{K}$, $u_L
= 2.35435 \times 10^{5}\,\mathrm{cm}\:\mathrm{s}^{-1}$ and $\rho_R =
1.24794 \times 10^{-12}\,\mathrm{g}\:\mathrm{cm}^{-3}$, $T_R =
207.757\,\mathrm{K}$, $u_R = 1.02987 \times
10^{5}\,\mathrm{cm}\:\mathrm{s}^{-1}$, where $L$ stands for the left
state, and $R$ the right state.  The initial discontinuity is at $x =
0$.  The gas is assumed to be ideal with an adiabatic index of $\gamma
= 5/3$ and a mean molecular weight of $\mu = 1$.  Initially, the
radiation is assumed to be in thermal equilibrium with the gas (i.e.,
$E_r = a T^4$).  The Planck and Rosseland coefficients are set to
$\kappa_{\mathrm{P}} = 3.92664 \times 10^{-5}\,\mathrm{cm}^{-1}$
and $\chi_{\mathrm{R}} = 0.848902\,\mathrm{cm}^{-1}$,
respectively.  The boundaries are held at fixed values.
Two refinement levels (three total levels) are used with the finest
resolution at $\Delta x \approx 2.9\,\mathrm{cm}$.  
In this test, a cell is tagged for refinement if the normalized second
derivative of either density or temperature given by \citep{flash}
\begin{equation}
  E_i = \frac{|u_{i+2} - 2 u_i + u_{i-2}|}{|u_{i+2}-u_i| +
    |u_i-u_{i-2}| + \epsilon (|u_{i+2}| + 2|u_i| + |u_{i-2}|)} \label{eq:error}
\end{equation}
is greater than 0.8.  Here, $\epsilon$ is a parameter set to 0.02 in
this test, and $u$ denotes either density or temperature.
The initial
conditions are set according to the pre-shock and post-shock states of
the M2 shock.  After a brief period of adjustment, the system settles
down to a steady shock.  The simulation is stopped at $t =
0.05\,\mathrm{s}$.  The results are shown in Figure~\ref{fig:M2}.  The
agreement with the analytic solution is excellent, except that the
numerical results in the figure had to be shifted by $10\,\mathrm{cm}$
in space to match the analytic shock position.  This discrepancy in shock
position is due to the initial transient phase as the initial state
relaxes to the correct steady-state profile.  No flux limiter is used
in this calculation 
because the analytic solution of \citet{LowrieEdwards08} assumes
$\lambda = 1/3$.

The setup of the Mach 5 shock problem is similar to that of the Mach 2
shock problem.  The computational domain in this test is
$-4000\,\mathrm{cm} < x < 2000\,\mathrm{cm}$.  Four refinement levels
(five total levels) are used with the finest resolution at $\Delta x
\approx 1.5\,\mathrm{cm}$.  The refinement criteria are the same as in
the Mach 2 shock test (Eq.~\ref{eq:error}).  The initial left and right states are
$\rho_L = 5.45887 \times 10^{-13}\,\mathrm{g}\:\mathrm{cm}^{-3}$, $T_L
= 100\,\mathrm{K}$, $u_L = 5.88588 \times
10^{5}\,\mathrm{cm}\:\mathrm{s}^{-1}$ and $\rho_R = 1.96405 \times
10^{-12}\,\mathrm{g}\:\mathrm{cm}^{-3}$, $T_R = 855.720\,\mathrm{K}$,
$u_R = 1.63592 \times 10^{5}\,\mathrm{cm}\:\mathrm{s}^{-1}$,
respectively.  Again, the system settles down to a steady shock after
a brief period of adjustment.  Figure~\ref{fig:M5} shows the results
at $t = 0.04\,\mathrm{s}$.  The results including the narrow spike in
temperature are in good agreement with the analytic solution.

\subsection{Advecting Radiation Pulse}
\label{test:pulse}

In this test, introduced by \citet{KrumholzKMB07}, we simulate the
advection of a radiation pulse by the motion of gas.  The initial
temperature and density profiles are
\begin{align}
  T = { } & T_0 + (T_1-T_0) \exp{\left(-\frac{x^2}{2w^2}\right)}, \label{eq:pulse}\\
  \rho = { } & \rho_0\frac{T_0}{T_1} + \frac{a \mu}{3R} \left(\frac{T_0^4}{T} - T^3\right),
\end{align}
where $T_0 = 10^7\,\mathrm{K}$, $T_1=2\times 10^7\,\mathrm{K}$,
$\rho_0=1.2\,\mathrm{g\:cm}^{-3}$, $w = 24\,\mathrm{cm}$, the mean
molecular weight of the gas is $\mu=2.33$, and $R$ is the ideal gas
constant.  Initially the radiation is assumed to be in thermal
equilibrium with the gas.  We also assume that the radiation pressure
is $E_r/3$ (i.e., $\lambda = 1/3$).  If there were no radiation
diffusion, the system would be in an equilibrium with the gas pressure
and radiation pressure balancing each other.  The interaction coefficients
are set proportional to density as $\kappa_{\mathrm{P}} = \chi_{\mathrm{R}}
= 100\,\mathrm{cm}^2\,\mathrm{g}^{-1} \times \rho$.  Because of
radiation diffusion, the balance is lost and the gas moves.  The
purpose of this test is to assess the ability of the code to handle
radiation being advected by gas.  We will solve the problem
numerically in different laboratory frames.  Then we can compare the
case in which the gas is initially at rest to the case in which the
gas initially moves at a constant velocity.

\begin{figure}
\plotone{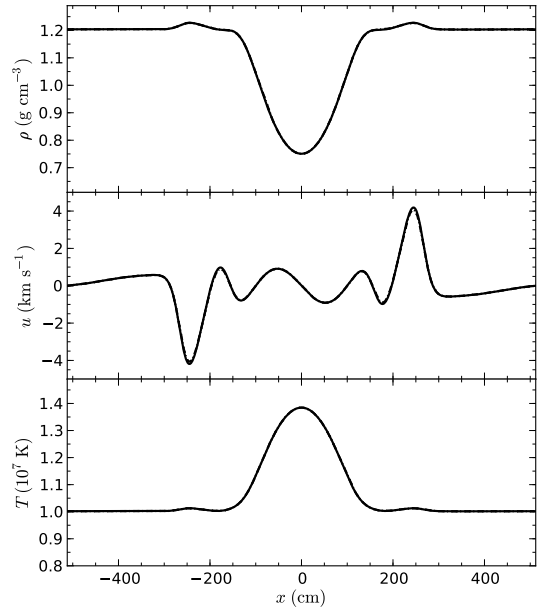}
\caption{ Density, velocity and temperature profiles at $t = 4.8
  \times 10^{-5}\,\mathrm{s}$ for the test of advecting radiation
  pulse.  The results of three runs are shown.  The difference is so
  small that it is nearly invisible to the eye.
  \label{fig:pulse}}
\end{figure}

\begin{figure}
\plotone{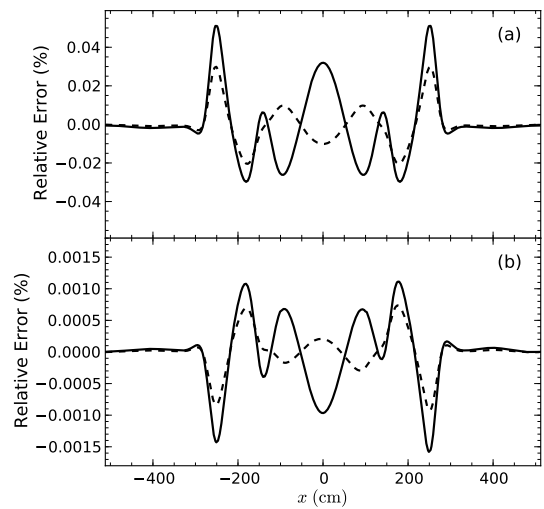}
\caption{ Relative errors in density ({\it solid lines}) and gas
  temperature ({\it dashed lines}) in the test of advecting radiation
  pulse.  The difference between the low-resolution unadvected and
  high-resolution unadvected runs is shown in panel (a), whereas the
  difference between the advected and unadvected runs in panel (b).
  \label{fig:pulse-error}}
\end{figure}

We have performed three runs for comparison with each other and
against the results of \citet{KrumholzKMB07}.  The computational
domain in all three runs is a one-dimensional region of
$-512\,\mathrm{cm} < x < 512\,\mathrm{cm}$ with periodic boundaries.
The simulations are stopped at $t = 4.8 \times 10^{-5}\,\mathrm{s}$.
The velocity in the first run is initially zero everywhere, whereas in
the second run it is $10^6\,\mathrm{cm\:s}^{-1}$ everywhere.  The
numerical grid in these two runs consists of 512 uniform cells.  The
third run is a high-resolution calculation of the first case with 4096
uniform cells.  Figure~\ref{fig:pulse} shows the density, velocity,
and temperature profiles for all three runs.  The results of the
advected run have been shifted in space for comparison.  The profiles
from these three runs are almost identical demonstrating the accuracy
of our scheme in this test.  We also show the relative difference in
Figure~\ref{fig:pulse-error}.  The relative error of the
low-resolution unadvected run with respect to the high-resolution run
is computed as (high - low)/high, and the relative error of the
low-resolution advected run with respect to the low-resolution
unadvected run is computed as (unadvected - advected) / unadvected.
Note that in Figure~\ref{fig:pulse-error} the results of the
high-resolution run have been restricted to the low-resolution grid by
averaging for comparison.  The figure shows that the difference
between the low-resolution and high-resolution runs is less than
0.05\% everywhere, and the difference between the advected and
unadvected runs is less than 0.0016\% everywhere.  The relative error
in our runs is 1000 times smaller than that of \citet{KrumholzKMB07}.
We also note that the asymmetry in the difference between the advected
and unadvected cases is expected because our upwinding scheme breaks
the symmetry slightly in the advected run.  The linear solver can
also break the symmetry slightly.

\subsection{Static Equilibrium}
\label{test:hse}

\begin{figure}
\plotone{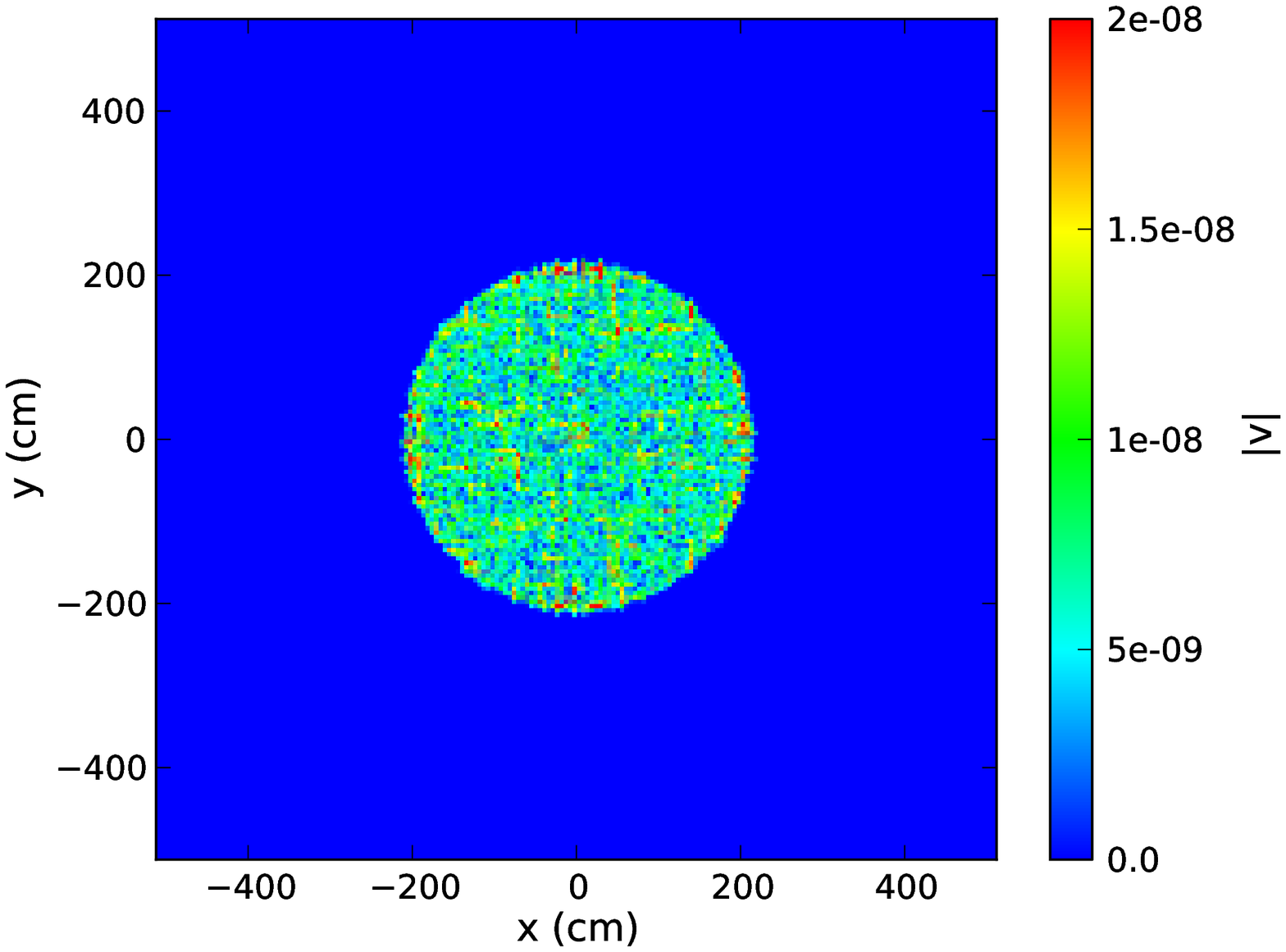}
\caption{ Structure of the magnitude of velocity at $t = 10^{-4}
  \,\mathrm{s}$ in the static equilibrium test.  The unit of velocity
  is $\mathrm{cm\:s}^{-1}$.
  \label{fig:pulse-hse}}
\end{figure}

In the test of advecting radiation pulse (\S~\ref{test:pulse}), if
there were no radiation diffusion, the system would be in a static
equilibrium with the gas and radiation pressures balancing each other.
In this test, we set both interaction coefficients to a very high value of
$10^{20}\,\mathrm{cm}^2\,\mathrm{g}^{-1} \times \rho$.  Thus almost no
diffusion can happen.  We have performed a calculation on a
two-dimensional Cartesian grid of $-512\,\mathrm{cm} < x <
512\,\mathrm{cm}$ and $-512\,\mathrm{cm} < y < 512\,\mathrm{cm}$ with
512 uniform cells in each direction.  The initial velocity is zero
everywhere.  For the initial setup, the coordinate $x$ in
Eq.~\ref{eq:pulse} is replaced by $r = \sqrt{x^2 + y^2}$.
Figure~\ref{fig:pulse-hse} shows the velocity profile at $t =
10^{-4} \,\mathrm{s}$.  The maximal velocity at that time is
$\sim 4 \times 10^{-8}\,\mathrm{cm\:s}^{-1}$. Note that the sound speed in this
problem is between $2.5 \times 10^{7}$ and $6.1 \times
10^{7}\,\mathrm{cm\:s}^{-1}$.  Such a small gas velocity indicates
that CASTRO can maintain a perfect static equilibrium in
multiple dimensions because of the way radiation pressure and gas
pressure are coupled in the Riemann solver.

\subsection{Radiative Blast Wave: Case 1}
\label{test:blast1}

In this test, a large amount of energy is deposited into a small
region.  This results in a spherical blast wave, which is somewhat
similar to the Sedov-Taylor blast wave in hydrodynamics, except here
we include radiation.  There is no analytic solution for this problem.

\begin{figure}
\plotone{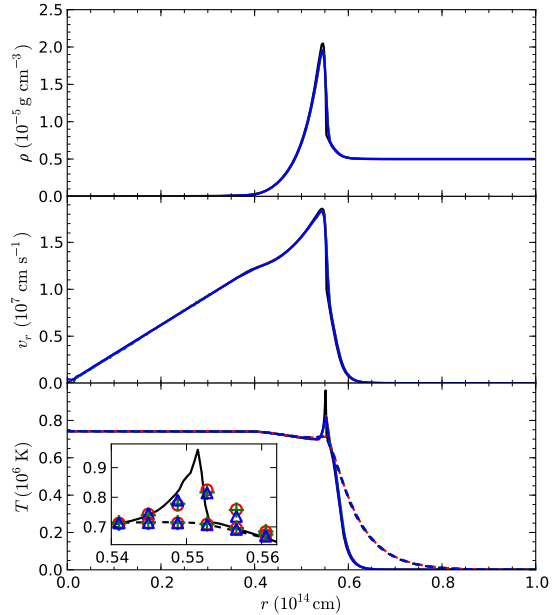}
\caption{ Density, radial velocity and temperature profiles at $t = 
  10^{6}\,\mathrm{s}$ for the first radiative blast wave problem.
  Here, radial velocity is the velocity in spherical radial direction.
  We show the results of the 1D spherical run ({\it red}), 2D cylindrical
  run ({\it green}), 3D Cartesian run ({\it blue}), and the
  high-resolution 1D spherical run ({\it black}).  The bottom
  panel shows both the gas temperature ({\it solid lines}) and
  radiation temperature ({\it dashed lines}).  The inset shows a
  blow-up of the spike in gas temperature, where the 1D, 2D, 3D, and
  the high-resolution runs are shown in red circles, green plus signs,
  blue triangles, and black lines.  
  \label{fig:blast}}
\end{figure}

\begin{figure*}
\epsscale{0.75}
\plotone{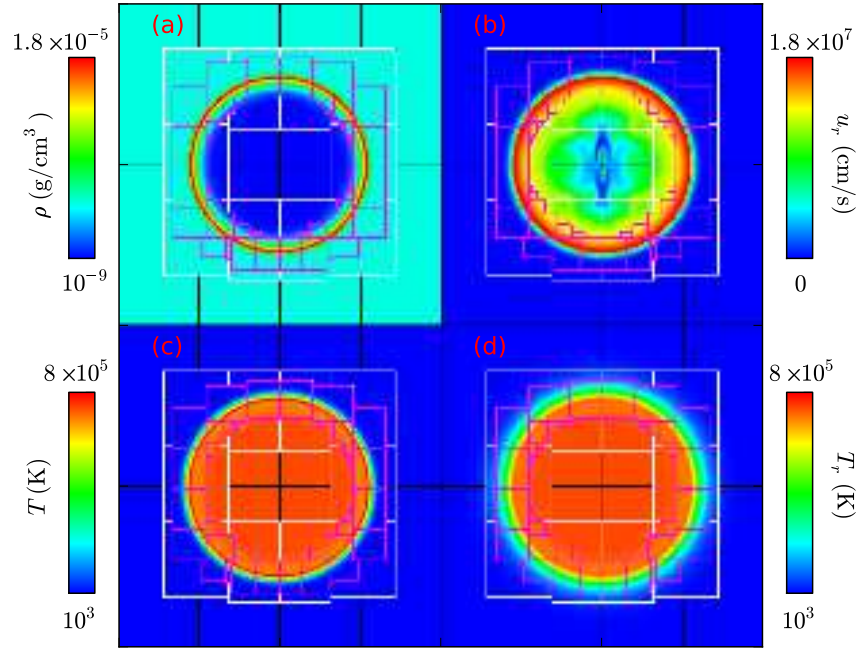}
\caption{ Snapshot of the 2D cylindrical simulation at $t =
  10^{6}\,\mathrm{s}$ for the first radiative blast wave test.
  Profiles of density, radial velocity, gas temperature, and radiation
  temperature are shown in panels (a), (b), (c), and (d),
  respectively.  Here radial velocity is the
  velocity in spherical radial direction, not the cylindrical radial
  direction.  The physical domain in each dimension is
  $(-10^{14},10^{14})\,\mathrm{cm}$ for each panel.  Also shown is the
  grid structure of the adaptive mesh.  Levels 0, 1, and 2 are shown
  in black, white, and purple, respectively. 
  \label{fig:bw2d}}
\end{figure*}

\begin{figure*}
\epsscale{0.75}
\plotone{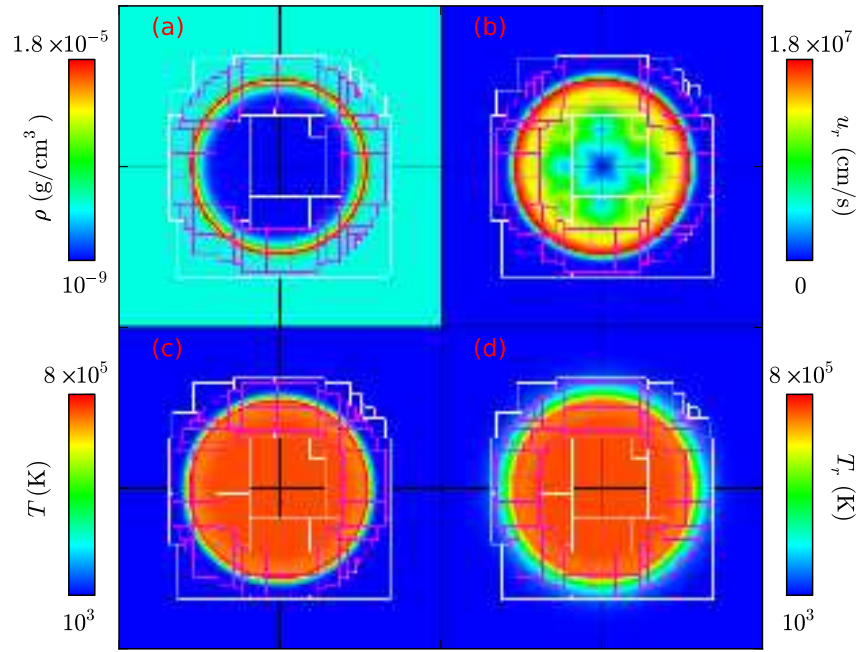}
\caption{ 2D slice at $z=0$ of the 3D Cartesian simulation at $t =
  10^{6}\,\mathrm{s}$ for the first radiative blast wave test.
  Profiles of density, radial velocity, gas temperature, and radiation
  temperature are shown in panels (a), (b), (c), and (d),
  respectively.  Here radial velocity is the
  velocity in spherical radial direction.  The physical domain in each
  dimension is 
  $(-10^{14},10^{14})\,\mathrm{cm}$ for each panel.  Also shown is the
  grid structure of the adaptive mesh.  Levels 0, 1, and 2 are shown
  in black, white, and purple, respectively. 
  \label{fig:bw3d}}
\end{figure*}

We use this problem to test the implementation of the geometric
factors in CASTRO.  We have performed simulations in 1D spherical, 2D
cylindrical ($r$ and $z$), and 3D Cartesian coordinates.  Initially,
the gas is at rest with a density of $5 \times
10^{-6}\,\mathrm{g\:cm}^{-3}$.  The initial temperature for both the
gas and radiation is set to $10^{3}\,\mathrm{K}$, except for a region
inside a sphere with a radius of $2 \times 10^{12}\,\mathrm{cm}$
centered at the origin where temperature is set to
$10^{7}\,\mathrm{K}$.  The gas is assumed to be ideal with an
adiabatic index of $\gamma = 5/3$ and a mean molecular weight of $\mu
= 1$.  The Planck and Rosseland coefficients are set to
$\kappa_{\mathrm{P}} = 2 \times 10^{-16}\,\mathrm{cm}^{-1}$ and
$\chi_{\mathrm{R}} = 2 \times 10^{-10}\,\mathrm{cm}^{-1}$.  These
simulations were run with two refinement levels (three total levels)
with a cell size of $\sim 3.9 \times 10^{11}\,\mathrm{cm}$ at the
finest level.  In these simulations, at each but the finest level a
numerical cell is tagged for refinement if it satisfies either $\rho >
5.01 \times 10^{-6}\,\mathrm{g\:cm}^{-3}$ or $T > 9 \times
10^6\,\mathrm{K}$.  Furthermore, the intermediate level is slightly
larger than the finest level due to proper nesting (see Paper I).  The
computational domain for the 1D spherical run is
$0<r<10^{14}\,{\mathrm{cm}}$.  The computational domain for the 2D
cylindrical run is $0<r<10^{14}\,{\mathrm{cm}}$ and
$-10^{14}\,{\mathrm{cm}}<z<10^{14}\,{\mathrm{cm}}$.  For the 3D run,
the computational domain is $(-10^{14},10^{14})\,{\mathrm{cm}}$ in
each direction.  A CFL number of 0.6 is used for these simulations,
and the initial time step is shrunk by a factor of 100 to allow the
point explosion to develop.  Note that we have chosen these initial
conditions so that different regimes can be explored by the test.  The
blast wave starts with almost all the energy in radiation.  At the end
of the simulations, approximately one third of the energy is in the
gas, and the gas pressure just behind the shock is about twice the
radiation pressure.  Furthermore, the gas and radiation are not in
equilibrium near the shock due to the low Planck mean opacity.

Since there is no analytic solution for the problem, we have also run
a high-resolution 1D simulation with a cell size of $\sim 4.9 \times
10^{10}\,\mathrm{cm}$ for comparison.  Figure~\ref{fig:blast} shows
the radial profiles of density, radial velocity, gas temperature, and
radiation temperature at $t = 10^6\,\mathrm{s}$ for these runs.  The
radial profiles of the 2D and 3D results are computed by mapping each
cell into its corresponding radial bin and averaging.  The width of
the bins is chosen to be the cell size at the finest refinement level.
The results from runs in three different coordinates are in excellent
agreement with each other, and they agree with those of the
high-resolution simulation.  A snapshot of the structure of density,
radial velocity, gas temperature, and radiation temperature at $t =
10^{6}\,\mathrm{s}$ for the 2D simulation is shown in
Figure~\ref{fig:bw2d}.  A 2D slice at $z=0$ of the 3D simulation is
shown in Figure~\ref{fig:bw3d}.  The multi-D results show good
agreement with each other, and they are spherically symmetric except
for some minor asymmetry in velocity at the low density region near
the center.  This asymmetry is, in part, due to the representation of
the initial hot sphere in non-spherical coordinates.  Another source
of the asymmetry in the 2D run is the coordinate singularity at the
longitudinal axis ($r=0$) of the cylindrical coordinates. It should
also be noted that the AMR gridding algorithm in CASTRO does not
necessarily preserve symmetry even if the layout of the numerical
cells that are tagged for refinement is symmetric.  Last, but perhaps
not least, the linear solver does not preserve perfect symmetry
either.

\subsection{Radiative Blast Wave: Case 2}
\label{test:blast2}

\begin{figure}
\plotone{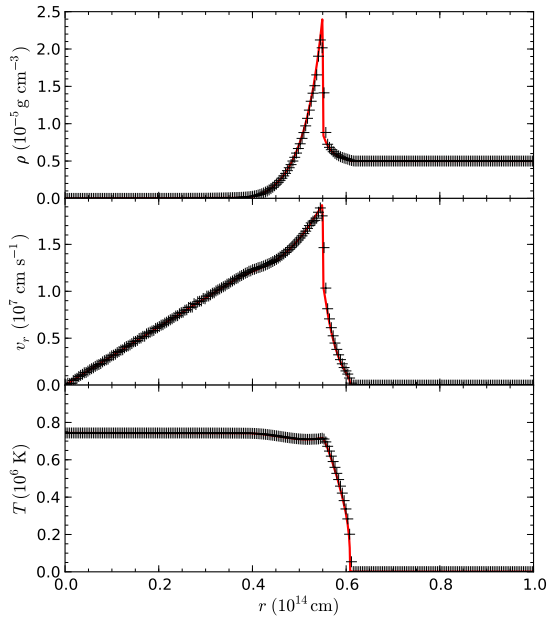}
\caption{ Density, radial velocity and temperature profiles at $t = 
  10^{6}\,\mathrm{s}$ for the second radiative blast wave problem.
  The results of the 2D cylindrical simulation are shown in symbols
  ({\it black plus signs}). 
  We do not show the radiation temperature because it is almost
  identical to the gas temperature.  Here, radial velocity is the
  velocity in spherical radial direction, not the cylindrical radial
  direction.  Also shown are the results of the high-resolution 1D spherical
  simulation ({\it red solid lines}).   
  \label{fig:blast2}}
\end{figure}

This test is similar to the first radiative blast wave test
(\S~\ref{test:blast1}) except that the Planck mean coefficient
is set to $\kappa_{\mathrm{P}} = 2 \times 10^{-7}\,\mathrm{cm}^{-1}$.
Thus the ratio of the two coefficients is
$\kappa_{\mathrm{P}} / \chi_{\mathrm{R}} = 1000$.  Because of such a
large ratio of the two mean opacities, the Lorentz transformation term
has to be treated implicitly in this test, otherwise the time step
would have to be much smaller for stability reasons.

We have performed a 2D cylindrical simulation with two refinement
levels (three total levels) and a cell size of $\sim 3.9 \times
10^{11}\,\mathrm{cm}$ at the finest level on a computational domain of
$0<r<10^{14}\,{\mathrm{cm}}$ and
$-10^{14}\,{\mathrm{cm}}<z<10^{14}\,{\mathrm{cm}}$.  In this 2D AMR
run, a numerical cell is tagged for refinement if it satisfies either
$\rho > 5.01 \times 10^{-6}\,\mathrm{g\:cm}^{-3}$ or $T > 9 \times
10^6\,\mathrm{K}$.  We have also performed a high-resolution 1D
simulation with a cell size of $\sim 4.9 \times 10^{10}\,\mathrm{cm}$
for comparison.  A CFL number of 0.6 is used for the simulations, and
the initial time step is shrunk by a factor of 100 to allow the point
explosion to develop.  Figure~\ref{fig:blast2} shows the radial
profiles of density, radial velocity, gas temperature and radiation
temperature at $t = 10^6\,\mathrm{s}$ for the two runs.  The results
show that CASTRO can handle a very large Lorentz transformation term
implicitly without having to decrease the time step.

\section{Summary}
\label{sec:sum}

We have developed a new radiation hydrodynamics solver in our
compressible astrophysics code, CASTRO.  We use the mixed-frame
approach and adopt the FLD assumption.  The solver uses a
second-order explicit Godunov method for the hyperbolic part of the
system and a first-order backward Euler method for the parabolic part.
We have also presented the mathematical characteristics of the
hyperbolic subsystem.  The eigenvalues and eigenvectors of the system
we have obtained are used to construct the Riemann solver in our
Godunov scheme, and could also be useful for other
characteristic-based schemes.

We have demonstrated the capability of CASTRO to address a wide range
of radiation hydrodynamics problems by extensive testing.  There are a
number of other radiation hydrodynamics codes.  Some recent examples
include ZEUS-2D \citep{TurnerStone01}, ZEUS-MP \citep{HayesNF06},
Orion \citep{KrumholzKMB07}, HERACLES \citep{GonzalezAH07}, V2D
\citep{SwestyMyra09}, Nike \citep{SekoraStone10}, RAMSES
\citep{CommerconTA11}, and CRASH \citep{crash}.  Here we compare
CASTRO with these other codes.
\begin{itemize}
\item A major advantage of CASTRO is its efficiency due to the use
  of AMR and combined with good scaling behavior on up to 32768 cores.  Among
  the other radiation hydrodynamics codes listed above, only
  Orion, RAMSES, and CRASH are three-dimensional AMR codes. ZEUS-MP
  and HERACLES are three-dimensional codes without AMR.  ZEUS-2D and
  V2D are two-dimensional codes.  Nike is a one-dimensional
  code.  
\item A unique strength of our code is that the hyperbolic solver in
  CASTRO is an unsplit version of the PPM method that avoids
  spurious noise caused by dimensional splitting (see Paper I for
  an example of the advantage of unsplit methods over dimensional
  splitting methods).  For the hyperbolic part, Orion, HERACLES,
  Nike, RAMSES, and CRASH use Godunov methods, whereas ZEUS-2D,
  ZEUS-MP, and V2D use the ZEUS type of algorithms.  The latter are
  faster but somewhat less accurate than high-order Godunov methods.
  However, for a radiation hydrodynamics code, the cost of the
  hyperbolic solver is no longer a concern, because the implicit
  parabolic solver is more expensive.  For multigroup radiation
  hydrodynamics, the implicit parabolic solver would be even more
  expensive and the hyperbolic solver would be essentially free.
\item Our scheme is very robust
  even for dynamic diffusion.  By solving the entire hyperbolic
  subsystem in one Riemann solver, the scheme can avoid the operator
  splitting errors that appear in many other numerical schemes
  \citep[see e.g.,][]{KrumholzKMB07}.  We note that RAMSES and CRASH
  also couple both radiation and matter in their Riemann solvers.
\item CASTRO is based on a mixed-frame formulation, as is Orion
  and Nike.  A main advantage of the mixed-frame approach is that it
  conserves the total energy, whereas its main disadvantage is that
  it is of limited use for line transport.  However, line transport
  cannot be treated by a gray radiation solver regardless of its
  choice of frame.  
\item CASTRO, ZEUS-2D, ZEUS-MP, Orion, V2D, RAMSES, and CRASH have
  adopted the FLD approach, whereas HERACLES and Nike are based on
  the two-moment approach.  Thus CASTRO carries the limitations of
  FLD, such as poor accuracy for optically-thin flows.  However, FLD
  is computational cheaper than the two-moment approach.  Moreover,
  for multigroup radiation, FLD is much less memory intensive than
  the two-moment approach.  
\end{itemize}

The current implementation uses a gray approximation based on the
frequency-integrated formulation of the radiation hydrodynamics
equations.  We note that numerical codes exist for multigroup
flux-limited radiation hydrodynamics
\citep[e.g.,][]{BurrowsLD07,SwestyMyra09,crash}.  It is
straightforward to extend our scheme to multigroup radiation because
the radiation pressure in the hyperbolic subsystem is still a
frequency-integrated quantity for multigroup radiation.  A multigroup
neutrino-radiation hydrodynamics solver is also currently under
development.

Further details on CASTRO can be found in the CASTRO User Guide
\citep{CASTROUserGuide}. 

\acknowledgments

The authors would like to thank Eric Myra, Doug Swesty and Michael Zingale
at Stony Brook University for a number of helpful discussions about
radiation hydrodynamics.
The work at LBNL was supported by the Office of High Energy Physics
and the Office of Advanced Scientific Computing Research
of the U.S. Department of Energy under contract
No.\ DE-AC02-05CH11231.
The work performed at LLNL was supported by
the SciDAC program of the U.S. Department of Energy under the auspices of
contract No.\ DE-AC52-07NA27344.
Adam Burrows was supported by the SciDAC program of DOE under grant number DE-FG02-08ER41544,
the NSF under subaward no. ND201387 to the Joint Institute for Nuclear Astrophysics, and
the NSF PetaApps program, under award OCI-0905046 via a subaward no. 44592 from Louisiana
State University to Princeton University.
This research used resources of the National Energy Research Scientific
Computing Center, which is supported by the Office of Science of the U.S.
Department of Energy under Contract No. DE-AC02-05CH11231.

\end{document}